\documentclass[12pt, preprint]{aastex}
\newcommand\ea{et al.\,}
\newcommand\chandra{{\it Chandra}}
\newcommand\xmm{{\it XMM-Newton}}
\newcommand\hst{{\it HST}}
\newcommand\eso{ESO 428-G14}
\newcommand\psc{\ifmmode{\rm\,cm^{-2}}\else{${\rm\,cm^{-2}}$}\fi}
\shortauthors{Levenson et al.}

\begin{document}
\title{Penetrating the Deep Cover of Compton Thick Active Galactic Nuclei}

\author{N. A. Levenson\altaffilmark{1}, 
T. M. Heckman\altaffilmark{2}, 
J. H. Krolik\altaffilmark{2},
K. A. Weaver\altaffilmark{2,3}, 
\& P. T. \.Zycki\altaffilmark{4}}
\altaffiltext{1}{Department of Physics and Astronomy, University of Kentucky,
Lexington, KY 40506; levenson@pa.uky.edu}
\altaffiltext{2}{Department of Physics and Astronomy, Bloomberg Center, Johns Hopkins University, Baltimore, MD 21218}
\altaffiltext{3}{Code 662, NASA/GSFC, Greenbelt, MD 20771}
\altaffiltext{4}{Nicolaus Copernicus Astronomical Center, Bartycka 18, 00-716 Warsaw, Poland}
\addtocounter{footnote}{4}

\begin{abstract}
We analyze observations 
obtained with the {\it Chandra X-ray Observatory}
of bright Compton thick
active galactic nuclei (AGNs), 
those
with column densities in excess of $1.5\times 10^{24}\psc$
along the lines of sight. 
We therefore view the  powerful central engines only indirectly,
even at  X-ray energies.  Using high spatial resolution and considering
only galaxies that do not contain circumnuclear starbursts, we
reveal the variety of emission AGNs alone may produce. 
Approximately 1\% of the continuum's intrinsic flux is detected
in reflection in each case.
The only hard X-ray feature is the prominent Fe K$\alpha$ fluorescence
line, with equivalent width greater than 1 keV in all sources.
The Fe line luminosity provides the best X-ray indicator of the
unseen intrinsic AGN luminosity.
In detail, the morphologies of 
the extended soft X-ray emission
and optical line emission  are similar, and
line emission dominates the soft X-ray spectra.
Thus, we attribute the soft X-ray emission to
material that the central engines photoionize.
Because the resulting spectra are complex and do not
reveal the AGNs directly, crude analysis techniques such
as hardness ratios would mis-classify these galaxies as
hosts of intrinsically weak, unabsorbed AGNs
and would fail to identify the luminous, absorbed nuclei
that are present.
We demonstrate that a three-band X-ray diagnostic can correctly
classify Compton thick AGNs, even when significant
soft X-ray line emission is present.
The active nuclei  produce most of the galaxies' total
observed emission
over a broad spectral range, and much of their light
emerges at far-infrared wavelengths.
Stellar contamination of the infrared emission can be
severe, however, making long-wavelength data 
alone unreliable indicators of the buried AGN luminosity.
\end{abstract}
\keywords{Galaxies: active --- galaxies: Seyfert --- X-rays: galaxies}

\section{Introduction}
Accretion onto supermassive black holes powers
active galactic nuclei (AGNs), but
intervening material 
prevents direct views of most sources.
The Compton thick  AGNs, which
have column density $N_H > 1.5 \times 10^{24} \psc$
along the line of sight to the
nucleus,  can be particularly
elusive, but
these  AGNs are  astrophysically 
important.  First, they are common, 
representing up to half of all Seyfert 2 galaxies
\citep*{Ris99}.
Second, in synthesis models \citep[e.g.,][]{Set89,Com95},
they are essential to replicate the observed spectrum of
the cosmic X-ray background, which peaks around 30 keV
\citep{Mar80}.

The intrinsic X-ray continuum of an AGN  up to about 100 keV 
may be
described by a power law, typically of photon index $\Gamma = 1.9$, or
spectral index $\alpha = 0.9$ \citep{Nan94}.
Intervening material preferentially
absorbs soft X-ray photons, and the direct emission of the central engine
is not detectable below 10 keV in Compton thick cases.  In the
obscuring medium, however, the intrinsic emission is strongly
reprocessed.  It emerges with an effective flat photon index and
diminished intensity, less than about 1\% of the intrinsic source
luminosity.
Thus,  Compton thick AGNs easily fall  below 
detection limits of current X-ray surveys, although the 
intrinsic power of their central engines is not exceptionally feeble.
These uncounted sources are responsible in part  for the diminishing fraction
of the X-ray background that is resolved toward higher energies
\citep{Wor05}.

In a Compton thick AGN, the large equivalent width (EW)
of the Fe K$\alpha$ fluorescence line is the most direct 
signature of its nuclear activity below 10 keV \citep*{Ghi94,Kro94}. 
The energy of this line is
6.4 keV when the fluorescing medium is not fully ionized
(less than \ion{Fe}{18}), or around 6.7 and 7.0 keV in
\ion{Fe}{25} and \ion{Fe}{26}, respectively. 
This line can also be an effective
diagnostic of the geometry of the obscuring medium
and the AGN's luminosity \citep{Lev02}.

With the powerful continuum suppressed at lower energies,
obscured AGNs reveal other
soft X-ray spectral components. 
The active nucleus may produce line emission
through photoionization and photoexcitation  
\citep{Sak00,Kin02,Ogl03}.
The AGN may also drive an outflow that generates
shocks in the surrounding medium, producing thermal emission.
Alternatively,  separate additional emission sources may
be present. 
Specifically, compact circumnuclear starbursts are common in active 
galaxies \citep*{Cid98,Gon01}.
and they  characteristically exhibit 
soft thermal X-ray spectra \citep{LWH01s}, similar to  starburst galaxies. 
Here we analyze seven high spatial resolution X-ray 
observations of  optically-bright, Compton thick Seyfert 2 galaxies
to identify X-ray and multiwavelength
signatures of their obscuration and 
to measure the power of their hidden central engines.

We define the sample in \S\ref{sec:sample}. 
We present the data and describe new results from the images in 
\S\ref{sec:obs}.   
With the spectroscopy results  (\S\ref{sec:spec}),
we use the Fe K lines to determine properties of the buried
AGNs, identify the soft X-ray emission as a consequence of 
photoionization, relate the soft X-rays to the continuum emission, and develop
coarse X-ray discriminants of 
Compton thick AGN (\S\ref{sec:results}).
We discuss the reflection geometry  of the continuum in \S\ref{sec:refl}.
We use infrared (IR) and optical data to
compare the
observable and intrinsic characteristics of these galaxies across the
electromagnetic spectrum (\S\ref{sec:optir}),
and we summarize our conclusions in \S\ref{sec:concl}.

\section{Sample Members\label{sec:sample}}
All sample 
galaxies are classified as normal  Seyfert 2s on the basis of
optical emission line  ratios. 
We require that the sample members be previously identified as Compton
thick and have bright AGNs, based on the core 
[\ion{O}{3}]$\lambda 5007$ line flux.
The latter criterion allows us to measure the nuclear
properties well.
In order to reveal the variety of X-ray emission for which an AGN alone
may be responsible,
we exclude galaxies that are known to contain nuclear starbursts.
Furthermore, to separate the immediate nuclear and circumnuclear 
environments, we require high spatial resolution, which only
\chandra{} can provide.
The exclusion of galaxies that contain starbursts and
the spatial resolution requirement render this
sample of Compton thick AGNs smaller than that of
other recent work \citep[such as ][]{Gua05}, but these restrictions
allow us to isolate the active nuclei and therefore 
recover their intrinsic properties accurately.
Data  are then available for seven galaxies, down to 
core [\ion{O}{3}] flux limit of 
$7\times 10^{-13} {\rm \, erg\,s^{-1}}$. 
We rely on Whittle's (1992) \nocite{Whi92} [\ion{O}{3}] fluxes,
where available.  
We use the  [\ion{O}{3}] fluxes of \eso, NGC 3393, and NGC 5347
from \citet*{Ber86}, \citet{Dia88}, and \citet{Gon96}, respectively. 
The sample is listed in Table \ref{tab:info}.  

\citet{Koy89} measured a large EW Fe line in a {\it Ginga} spectrum 
of NGC 1068 and argued that the direct AGN continuum must therefore
be completely blocked. 
\citet{Mai98} identified \eso, NGC 1386, and NGC 3393 as Compton thick 
in {\it BeppoSAX} observations.
In the first, they noted the flat ($\Gamma = 0.57$) spectrum, 
low ratio of observed 2--10 keV to [\ion{O}{3}]$\lambda 5007$
flux, and non-detection above 15 keV as indirect evidence for 
$N_H > 10^{25} \psc$.
In the second,  they fit the 1--10 keV spectrum with a reflection
model and a prominent Fe K$\alpha$ line.
They fit
the 1--80 keV  spectrum of NGC 3393 to find $N_H > 10^{25} \psc$.
\citet{Ris99} identified NGC 5347 as Compton thick on the basis of
the large Fe $K\alpha$ EW measured in a spectrum from {\it ASCA}
and the low ratio of observed X-ray/[\ion{O}{3}] flux.
\citet{Ris00} suggested that NGC 7212 is Compton thick
from analysis of a low signal-to-noise {\it ASCA} spectrum,
finding an apparently flat continuum and a prominent Fe line.
\citet*{Gua05b}  confirm this identification in
a higher quality spectrum obtained with \xmm.
\citet{Awa91} first measured the large column density and prominent
Fe K line of
Mrk 3 in a spectrum they obtained using {\it Ginga}.

Population synthesis analysis of optical spectra 
shows intermediate-aged
and older stellar populations (ages $> 100$ Gyr) in the nuclei of nearly
all sample
members \citep[and references therein]{Cid01,Cid04}.
The exception, NGC 5347,  exhibits
no significant emission from polycyclic aromatic hydrocarbons
at 3.3$\mu$m \citep{Ima03} and
far-infrared (FIR) colors that are characteristic of AGNs
\citep*{deG87}.
Thus, we do not expect any significant stellar contributions to
the X-ray spectra of these galaxies.

\section{Observations and Image Analysis\label{sec:obs}}

We observed the 
galaxies using
the \chandra{} Advanced CCD Imaging Spectrometer (ACIS)
back-illuminated S3 
detector. 
We reprocessed all data from original Level 1 event files using Chandra Interactive
Analysis of Observations (CIAO) software, version 3.2.
(See the \chandra{} Science Center\footnote{http://cxc.harvard.edu/} 
 for details about \chandra{}
data and standard processing procedures.)
We applied current calibrations (Calibration Database version 3.0), 
including corrections for
charge-transfer inefficiency and time-dependent gain variations, to produce
the Level 2 event files.
In each case, we included only good events that 
do not lie on node boundaries, where discrimination of cosmic rays
is difficult.
We examined the lightcurves of background regions and excluded
times of significant background flares.

In the NGC 1068 \dataset [ADS/Sa.CXO#obs/00370]{observation},
we use only the data with 0.1 s ``frame times.''
With the CCDs read out after very short exposures, we
minimize  the problem  of
``pileup,'' in which coincident photons that arrive within
a single CCD readout are not measured accurately.
We otherwise applied the same data reduction procedures described above.
This particular observation and observing mode required
a correction to the on-source 
exposure\footnote{http://cxc.harvard.edu/ciao/caveats/acis\_interleave.html},
for a net exposure of 1.53 ks.
Mrk 3 was observed with the High-Energy Transmission
Grating in place.  We therefore do not analyze the extended
emission of this galaxy, and we refer to published measurements of the spectrum 
\citep{Sak00} where comparisons are appropriate.

Table \ref{tab:info} lists observation dates and resulting exposure times
of the data we analyze directly 
along with some basic information about all of these galaxies.
The Galactic column density toward each source ($N_H$; column 4) 
is based on the \citet{Schl98} measurements of $A_V$, 
with $N_H = 1.9\times 10^{21} A_V {\rm \, cm^{-2} \, mag^{-1}}$.
We use $H_0=70 {\rm \, km\,s^{-1}\,Mpc^{-1}}$ and  redshift ($z$; column 5) 
to derive distances (column 6) and physical
scales (column 7).

\begin{figure}
\centerline{\includegraphics[width=6in]{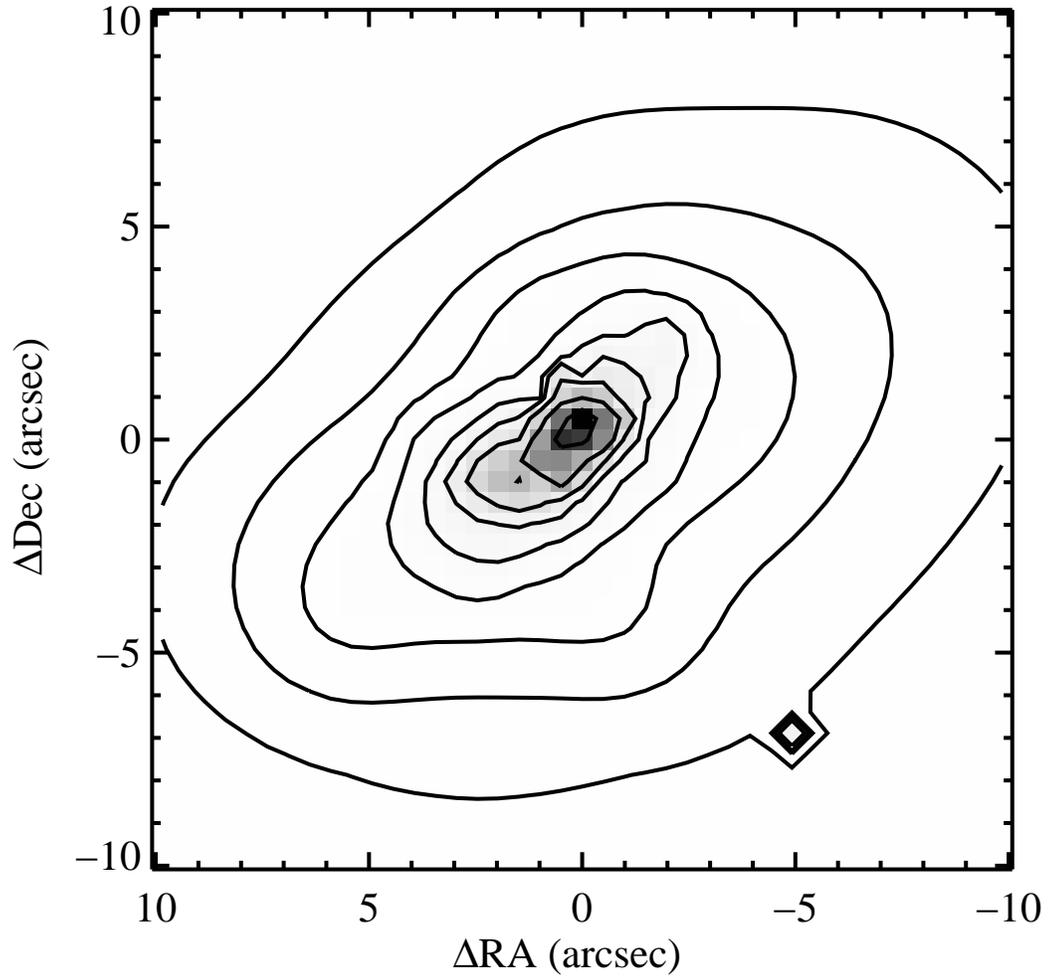}}
\caption{\label{fig:imgs} 
X-ray image of \eso.
The image
has been adaptively smoothed and is scaled linearly.
The contours are spaced logarithmically.
}
\end{figure}

The \chandra{} image of \eso{} (Figure \ref{fig:imgs}) is typical.
The images clearly show extended X-ray emission, demonstrating
that the unresolvable central engine is not the sole observed X-ray source.
The resolved emission extends over scales of hundreds of parsecs, and it
is predominantly soft.  
We compare images at 
soft (0.3--1 keV), medium (1--4 keV), and hard (4--8 keV) energies
with  corresponding models of \chandra's point spread
function. 
\eso, NGC 1068, NGC 1386, and NGC 3393 are significantly extended in the soft and
medium bands, even on small ($1\arcsec$) scales, 
while their hard bands are unresolved.
NGC 7212 is significantly extended 
only in the medium band. 
The bright center of  NGC 5347 is not resolved in any band,
although this galaxy shows distinct  emission 
approximately 3\arcsec{} to the northeast  of the nucleus.

\begin{figure}
\centerline{\includegraphics[width=6in]{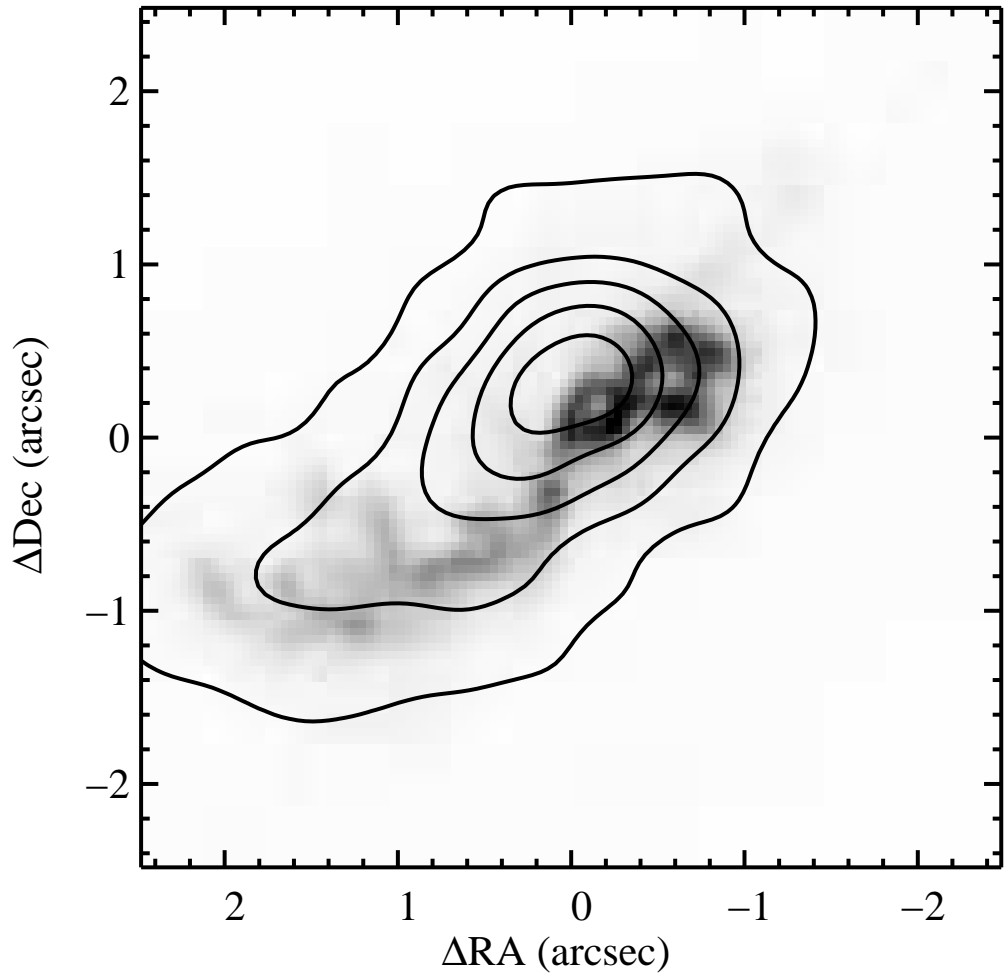}}
\caption{\label{fig:hst} 
Continuum-subtracted H$\alpha$ image of \eso{}
with linearly-scaled X-ray contours overlaid.
The similar morphology of the X-ray and optical line emission suggests
their common photoionized origin in all
the sample members.
}
\end{figure}

In all cases, the large-scale extended X-ray emission 
appears similar to images of these galaxies in optical emission lines
(Figure \ref{fig:hst}), as \citet*{Bia06} have shown for some of these sample members.
For example, the morphology of  \eso{} in X-rays specifically and in detail resembles that
of both H$\alpha$  and [\ion{O}{3}] \citep{Fal96}. From the nucleus (evident at both radio and
X-ray energies), 
very bright emission extends
toward the northwest, with fainter emission extending toward the southeast.
In Figure \ref{fig:hst}, we show the \chandra{} intensity contours
overlaid on the  continuum-subtracted H$\alpha$
image obtained with 
the {\it Hubble Space Telescope} (\hst) 
Planetary Camera of the  Wide Field and Planetary Camera 2 (WFPC2).
We aligned and scaled the image  through the 
 F814W filter (\dataset[ADS/Sa.HST#U2NP0603T]{U2NP0603T}) 
to subtract continuum from the F658N image (\dataset[ADS/Sa.HST#U2NP0601T]{U2NP0601T}). 
We then aligned the X-ray-identified nucleus (\S \ref{sec:spec})
with the maximum optical continuum emission.  
The maximum broad-band X-ray emission is located 
north of the Fe K$\alpha$ emission peak that defines the X-ray center.
Thus, the highest-intensity contours are not centered on the optical nucleus.
Photoionization by the central engine produces the optical line emission 
in all these galaxies.  
While the lower spatial resolution X-ray observations 
cannot show variations on the very small scales
($0\farcs1$) that are evident in the 
optical data,
the similar overall X-ray morphology suggests that 
photoionization is also the origin of the soft X-ray line emission.
The \chandra{} spectra also support this interpretation of 
the line emission as a consequence of 
radiative processes rather than collisional excitation as
(\S\ref{subsec:photo}). 

We find several sources in addition to the active nucleus 
near each galaxy's center.  We present these results in Appendix \ref{app:ptsrc}.

\section{Spectroscopy\label{sec:spec}}
The ACIS data afford spatially resolved spectroscopy
of several interesting regions.
We obtained
spectra of the nuclei, extended circumnuclear regions, and 
the whole galaxies.
We identified the locations of the AGNs in 
continuum-subtracted Fe K$\alpha$ images.
Line emission dominates the emission
in images extracted with energy restricted $6 < E <7 $ keV.
We measure the corresponding continuum in the bandpass $4.5 < E < 5.5 $ keV,
at which the ACIS response is similar.
We smoothed each of these images by a Gaussian of FWHM $ = 1\arcsec$, then
subtracted the continuum (4.5--5.5 keV) image from the corresponding line+continuum 
(6--7 keV) image without any further scaling.
This procedure is adequate to determine the locations of the maximum Fe K$\alpha$ emission,
but we do not use these images for quantitative measurements of the line flux. 
The hard X-ray peaks measured in the continuum-subtracted images are listed in Table \ref{tab:info}.

To isolate the AGN emission, we 
extracted spectra within an aperture
of $1\farcs5$ radius centered on each of these nuclear locations.
Even in nearby galaxies, however, these unresolved ``nuclear'' apertures
cover hundreds of parsecs around the active nuclei.
While spectroscopy of large regions of some of these sample members
has been presented elsewhere 
(\citealt*{You01}; \citealt{Bia06}), 
the small nuclear apertures
are critical in order to measure the immediate AGN emission, especially the
Fe K$\alpha$ line flux and EW. 
In all cases, we subtracted the background measured in nearby areas 
that contain no obvious sources.
Over the course of the \chandra{} mission, 
the soft X-ray sensitivity has diminished, likely the
result of build-up of material on the detector.
We created ancillary response files that account for
this time-varying effect.
We performed the model fitting in XSPEC \citep{Arn96}.

Broadly, the nuclear spectra are similar.  
The hard X-ray emission ($E> 3$ keV) is spectrally flat and relatively
weak, with a prominent Fe K$\alpha$ emission line in each case.
The bright soft X-ray spectra show line emission.
In order to measure the Fe K line accurately and to isolate the 
AGN continuum, we first fit the spectra from only 4--8 keV.
We fit unbinned spectra using the $C$ statistic \citep{Cas79} to
retain significant counts at energies greater than 6.5 keV.
We model the Fe line as an unresolved Gaussian, accounting for
each galaxy's redshift in measuring the central energy of the line.

The observed flat hard X-ray spectra and previous  evidence for 
Compton thick obscuration
indicate that we cannot directly detect the intrinsic AGN power law
continuum in these cases.  
Instead, we model the observed continuum as purely reflected AGN
light in a neutral medium using the PEXRAV model \citep{Mag95} in XSPEC.
Within the \chandra{} bandpass, we are not sensitive to some
of the model parameters (e.g., intrinsic photon index and high-energy cutoff),
and others are degenerate. 
Thus, we fit only for normalization of the intrinsic power law,
fixing the remainder of the model parameters:
intrinsic photon index $\Gamma = 1.9$, high-energy cutoff $= 300$ keV, and
abundances are solar \citep{And89}.
The model describes reflection off a 
plane-parallel, semi-infinite slab, and it is a function
of viewing angle, measured from the slab normal.
We adopt inclination $i = 63\degr$, the model default.
This is an intermediate value considering 
obscured views only; assuming a toroidal obscuring
geometry with half-opening angle $\theta$, 
the central engine is not viewed directly from $i=\theta$ to 90\degr. 
Although the high-energy ($E > 10$ keV) spectral shape varies with viewing 
angle, the primary effect on the \chandra{} spectra is to
alter the normalization.
Compared with the extreme values,
the adopted inclination angle results in 
normalizations that may be overestimated by up to 25\% 
(for true pole-on views) or underestimated by up to
a factor of three (for equatorial views).
We also account for Galactic absorption, although
its effect is negligible.

\begin{figure}
\centerline{\includegraphics[width=6in]{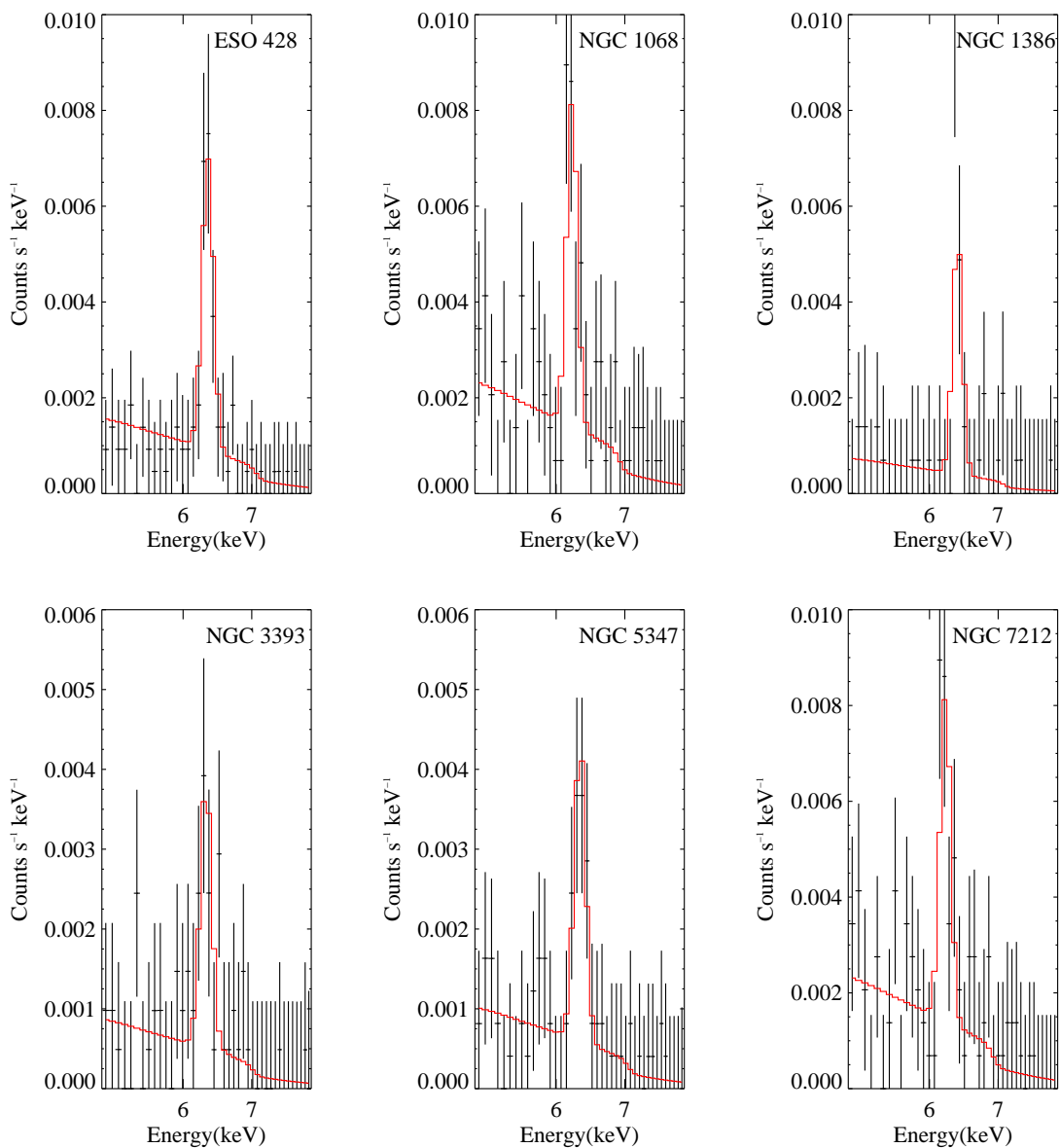}}
\caption{\label{fig:fespec} 
The prominent Fe K$\alpha$ lines are evident in all the observed nuclear
spectra.  The best-fitting  model of the hard emission, 
which includes a pure
reflection continuum and the emission line, is plotted with the data. 
We show binned data here but use the unbinned data in the model fitting.
}
\end{figure}

Table \ref{tab:hard} contains the results of these hard nuclear fits, including
the $C$ statistic and number of degrees of freedom ($dof$). 
The tabulated values of the 
power law normalization (column 5)
assume full (100\%) reflection of the AGN continuum.
With decreasing relative reflection, (e.g., with 
smaller covering fraction of reflecting material),
the underlying AGN normalization increases proportionally.
Figure \ref{fig:fespec} shows the hard spectral fits; the
prominent Fe lines are evident in all these nuclear spectra.

At the current sensitivity and resolution, there are no significant
spectral differences
among an intrinsic (heavily absorbed) continuum, this reflection model, 
and an unabsorbed $\Gamma = 0 $ power law.
For example, the Fe absorption edge is strongest in the first of these 
models and absent in the last,
but the edge cannot be distinguished in any of these data.
The shortcoming of the flat power law is that this model 
is purely phenomenological;
such intrinsic spectra are not observed
in Seyfert 1 galaxies or less-obscured Seyfert 2s, 
where they would be detectable.
The advantage of adopting the  reflection model is that
it does provide a lower limit on
the intrinsic AGN power,
which we will compare with other multiwavelength observations
and calculations (\S\ref{sec:optir}). 

The soft X-ray spectra do not affect the high-energy results, so
for simplicity, we fixed these latter model components while 
fitting the spectra over the full 0.4--8 keV range.
We grouped the full spectra  into bins having a minimum
of 20 counts each and use the $\chi^2$ statistic.
While fixing foreground extinction at the Galactic
values, we considered several physically-motivated
model components to account for the soft X-rays:
the MEKAL thermal model  \citep*{Mew85,Mew86,Lie95}, 
additional power laws, and line emission.
An AGN-driven outflow may generate thermal emission.
The non-thermal continuum may be the intrinsic continuum
reflected off the surrounding medium.
Photoionization by
the AGN continuum may produce recombination and fluorescence lines.

\begin{figure}
\centerline{\includegraphics[width=5in]{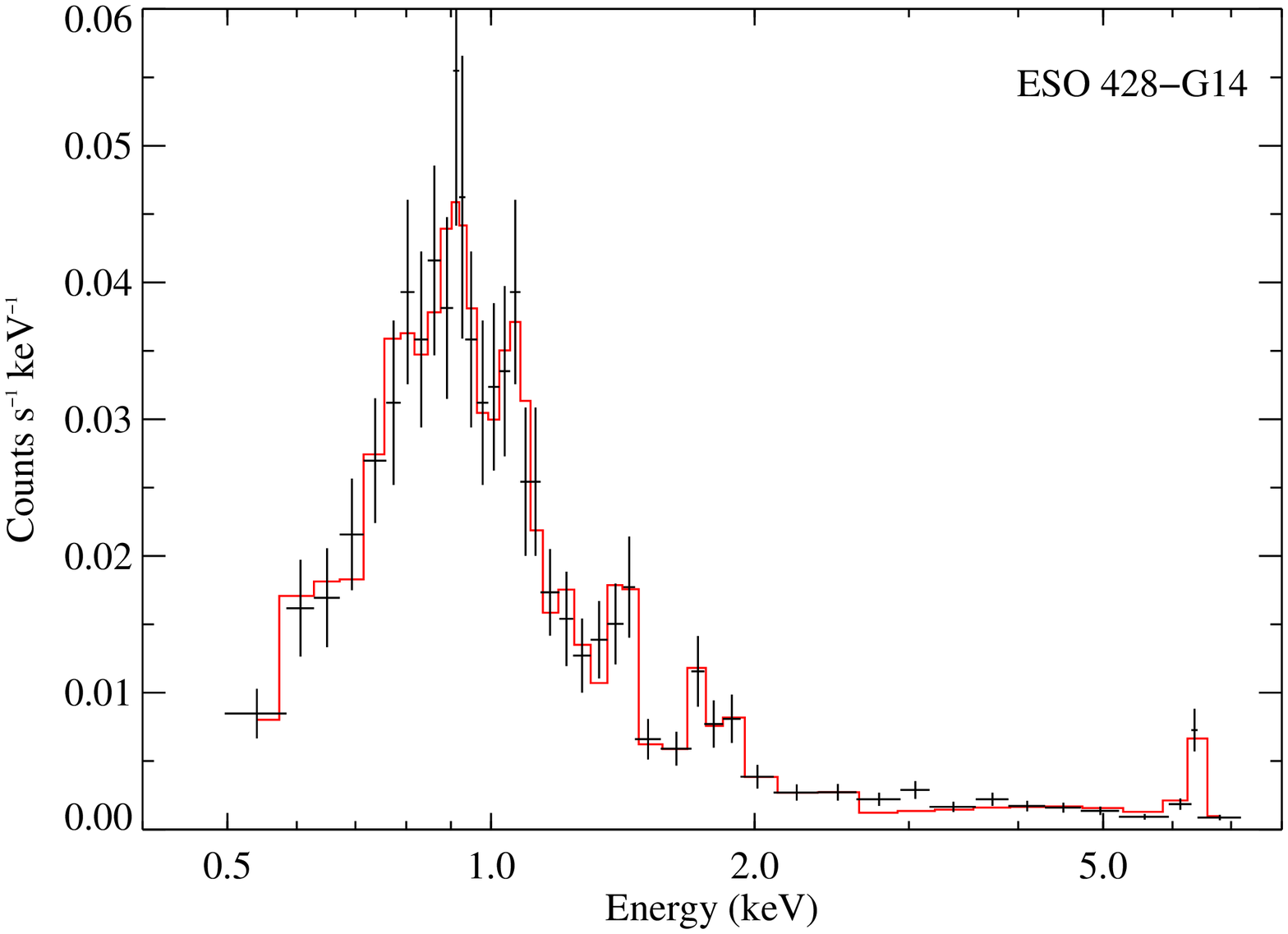}}
\centerline{\includegraphics[width=5in]{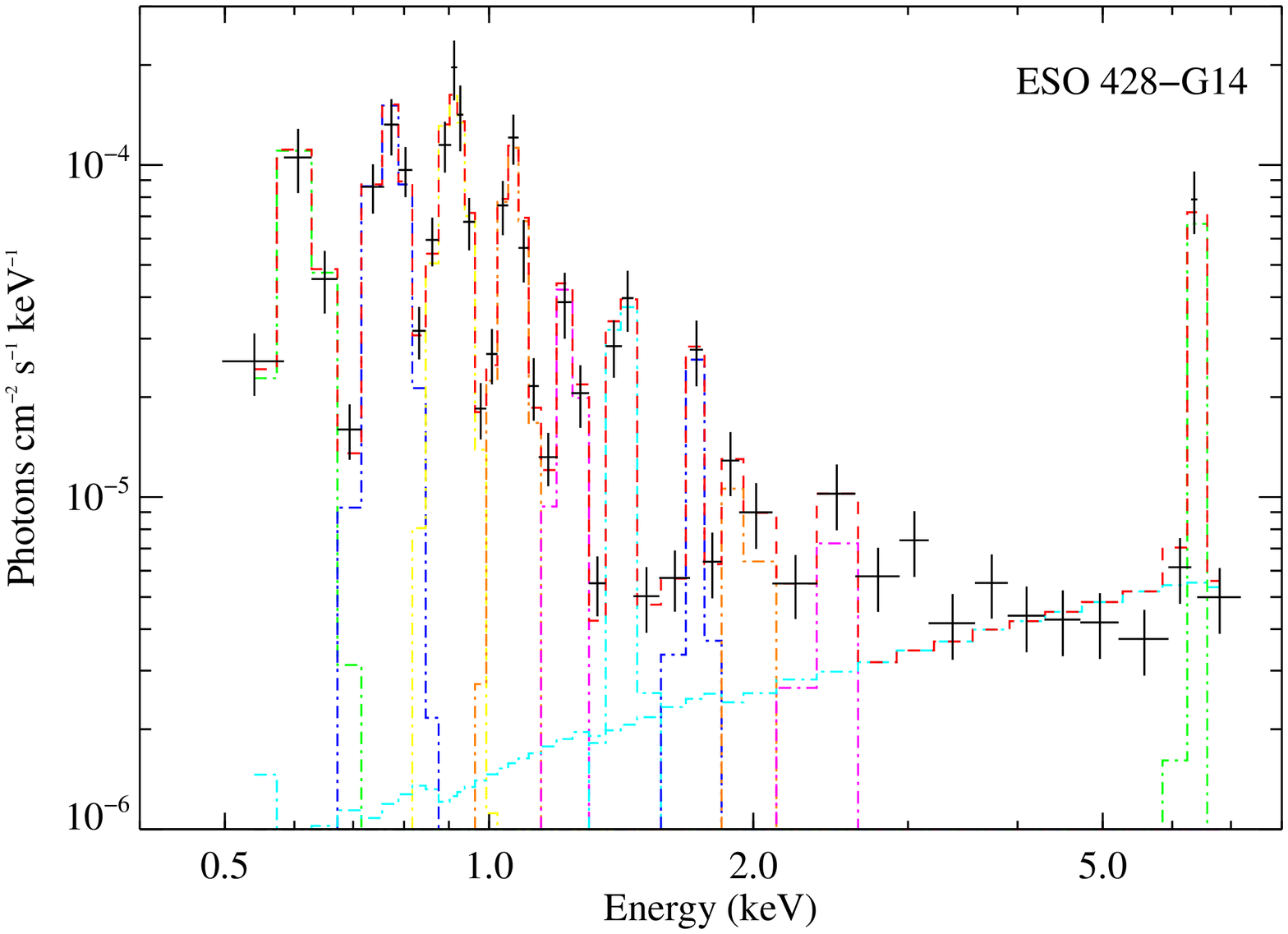}}
\caption{\label{fig:speceso} 
Nuclear spectrum of \eso{} ({\it crosses}) and 
best-fitting model ({\it red histogram}).
({\it upper panel}) Observed spectrum and total model, uncorrected
for detector sensitivity.
({\it lower panel}) Emitted spectrum and model, corrected
for detector sensitivity. Individual model components are plotted
with dot-dashed lines.
The hard X-ray spectrum shows
the reflected AGN continuum and the Fe K$\alpha$
line, while 9 unresolved Gaussian line profiles
account for the soft X-rays.
}
\end{figure}

Overall, we find that the line emission dominates the
nuclear spectra below 3 keV.
While the ACIS detector cannot resolve individual transitions,
the spectra show relatively sharp emission peaks, as opposed to
the much broader blends of many lines that 
collisional excitation would exhibit.
We model these lines as unresolved Gaussians.  In each
spectrum, we include additional lines or other
components as long as they are
significant at the 95\% confidence level, based on an $F$ test.
Figure \ref{fig:speceso} of \eso{} is representative of
the full nuclear spectra and the best-fitting models.
We emphasize that predominantly thermal models do not
quantitatively fit any of 
the observed spectra as well as these line models do 
(\S\ref{subsec:photo}).

We consider the spectra of extended regions of the galaxies.
(We exclude NGC 1068 from this discussion because its
extended X-ray emission has been extensively analyzed previously; \citealt{You01}.)
We  extracted spectra from the bright, central regions of the
galaxies, 
including their nuclei.  These elliptical apertures 
are   $6\arcsec \times 3\arcsec$ in NGC 1386 and NGC 3393,
and 
$5\arcsec \times 3\farcs5$ and $ 3\farcs5 \times 2\arcsec$
in \eso{} and NGC 7212, respectively.
We also obtained a spectrum of the soft X-ray emission northeast of the nucleus of
NGC 5347, within a circular aperture of $1\farcs5$ radius.
The total emission  within $1\arcmin$ of each nucleus
yields a spectrum  of each galaxy as a whole.

Tables \ref{tab:eso} and \ref{tab:n3} list  
the results of these model fits of the soft nuclear spectra and
the larger central regions where our findings are new or significantly
different from published values, and Table \ref{tab:flux} summarizes
the flux measurements. 
Some of the spectra do require model components in
addition to the unresolved lines and hard X-ray features.
The central \eso{} spectrum requires an additional continuum
source, which a power law describes well.  Because the
power law index is not well constrained, however, we
fix its slope at 1.9, matching the intrinsic AGN
continuum.

Given the limited spectral resolution of the detector,
each modeled line in the soft spectrum represents a blend of emission lines
and is not attributable to a single transition.
However,  we identify a particular element as the
dominant emitter in several cases 
where  the observed line energy corresponds to 
a known (measured or theoretical) value.
For example, in all spectra we find emission near
0.56  and 0.91 keV, which we identify as  
\ion{O}{7} and \ion{Ne}{9}, respectively.
Several of the spectra suggest \ion{Si}{13} near 1.8 keV.
We do not characteristically find emission from fully-ionized species,
although we measure \ion{O}{8} in NGC 1386 and it may appear
blended in the spectrum of \eso.
 \ion{Ne}{10} may also contribute  to the lines observed at 
1.0 keV in several of the spectra.
Most of the lines observed to have energies between
0.7 and 1.2 keV likely contain a strong contribution 
from the many Fe L shell lines in this energy range.

The best-fitting models of the NGC 3393 nuclear and central
spectra each include a thermal component.  At this temperature,
($kT \approx 1.6$ keV), the collisional excitation results
in a number of emission lines around 1 keV,  where Fe L 
transitions are also strong.  We considered 
adding to the unresolved lines  a broad
Gaussian emission feature, to account for the many Fe L
lines, instead of the thermal component.
The model with the thermal component fits the data much better
than this alternate purely descriptive model.
In the nuclear spectrum, for example,
$\chi^2/\nu = 25/21$ for Gaussian width $\sigma = 0.3$
and central energy 0.9 keV.  We note that both our hard and soft
spectral fits differ from the results of \citet{Bia06}.
The larger Fe K$\alpha$ EW we find may be a consequence of our smaller 
aperture, selected to isolate the line emission against the AGN's continuum.
We do not, however, significantly measure as many unresolved soft lines 
as they report,
even in our larger aperture. 

The emission northeast of the NGC 5347 nucleus is distinct, but faint.
Only the soft X-ray flux is significant.  
A single-temperature thermal model fits the unbinned spectrum of this region,
with $kT = 0.24 \pm 0.05$, although optical line ratios
indicate that this emission is photoionized \citep{Gon96}.

We fit the whole galaxy spectra with models based on the best-fitting
nuclear and central models, freeing the both the hard and soft X-ray
parameters.  These spectra cover very large areas, so they contain the
integrated flux of many different emission sources.  We do not
interpret the physical significance of the models in detail, but use
these results for flux comparisons (Table \ref{tab:flux}).  

Most of
the total flux is located within the smaller (central and nuclear)
regions, rather than extended throughout the host galaxies.  
The central regions 
account for more than 70\% of each
galaxy's total soft emission.  
Typically, the nuclear aperture alone 
contains 
most of the soft emission.  
NGC 3393 is exceptional;
here less than 40\% of the
soft flux is contained within the nucleus, 
and the larger ($1700 \times 770$ pc)
central area accounts for the plurality of this galaxy's soft emission.
In each case, 
the nucleus alone does provide at least half of the total observed galactic hard X-ray emission
(measured within a $1\arcmin$ radius aperture).
In most cases, 
however, the {\em extra}-nuclear fraction
of observed hard X-rays is significant, around 40\% to 50\%. 
In a less-obscured active galaxy, nearly all of the
hard X-rays would be detected in the immediate vicinity
of the unresolvable central engine, because the hard AGN continuum would be observed directly.
In this group of Compton thick sources,
the spectra show that the hard X-rays emerge indirectly.
As a result, line emission at energies greater than 2 keV is significant 
and may arise on larger physical scales.
Only in NGC 5347 is the net (line and continuum) reprocessing confined to a small 
region (radius $<$ 300 pc).

We estimate that  X-ray binaries contribute a small fraction 
of the observed total hard X-ray flux of the whole galaxies we measured.
\citet{Col04} derive scaling relations for the 0.3--8 keV flux
of unresolved sources in terms of galaxy luminosity measured at 
FIR, $K$, and UV wavelengths.
To determine the UV luminosity, we use {\it IUE} observations 
of NGC 3393 \citep{Kin93}. In the other cases, we very roughly
estimate this
value from observations at $B$, assuming 
$f_{2000\AA}/f_{4400\AA} = 0.2$ \citep{Kin96}.
We find that the 0.3--8 keV flux of X-ray binaries is
less than about 10\% of the total 2--10 keV flux.  Modeling the
typical source as a power-law
($\Gamma=1.8$) spectrum with Galactic absorption, we conclude
that the 2--10 keV flux of such point sources is less than
5\% of the total 2--10 keV flux of each galaxy. 

\section{Results\label{sec:results}}
\subsection{Fe K$\alpha$ Line Emission: An Indicator of the Intrinsic Nuclear Luminosity\label{subsec:feka}}

The line emission near 6.4 keV is the most significant X-ray 
characteristic of the AGNs in these galaxies.  All of the
line central energies are consistent with Fe K$\alpha$ in a 
``neutral'' medium, i.e., less ionized than \ion{Fe}{18}.  
To remain neutral, 
the fluorescing medium must not receive the powerful ionizing
flux of the central engine directly, and it is
likely the torus
of the unified AGN scenario \citep{Ant93}, the same region
that blocks broad optical emission lines from direct view in
these Seyfert 2s.

The line EW is a function of obscuring column density, covering
fraction, and viewing angle \citep{Kro94}.
Qualitatively, larger covering fraction produces larger EW, since more
Fe-edge photons are captured and generate K$\alpha$ photons.  
The continuum is most suppressed along
the equatorial plane, so these viewing angles also result in larger EWs.
The large equivalent widths 
confirm the Compton thick identification of these AGNs.
At solar abundance, EWs larger than 1 keV require $N_H > 1.5\times 10^{24} \psc$. 

The EWs place some limits on the covering fraction, dependent on the viewing
angle, $i$ \citep{Kro94,Lev02}.
In these models, the intrinsic AGN spectrum has $\Gamma = 1.9$, 
and the obscuring torus has a square cross-section
allowing for unobstructed views over half-opening angle $\theta$.
An EW $< 1.5$ keV does not strongly constrain the covering fraction,
yielding $\theta \lesssim 50\degr$.
With the large EWs of \eso{} and NGC 1386, we find
$\theta < 35$ and $20\degr$ for modest viewing angle ($i= 65\degr$),
and 
$\theta < 45$ and $30\degr$, respectively, viewed almost
through the equatorial plane ($i = 85\degr$).

The results of these numerical simulations also 
show that the Fe line luminosity and the intrinsic AGN luminosity
are related.  For each of the observed AGNs, 
we use these numerical models to find the region in inclination angle, 
covering fraction,
and column density parameter space consistent with the measured EW
and its errors.   
We then
compute the mean ratio of Fe K$\alpha$ to intrinsic continuum luminosity 
within this
portion of parameter space.
We typically find $L_{Fe}/L_{AGN,2-10} \approx 2\times 10^{-3}$,
where $L_{AGN,2-10}$ is the  AGN's intrinsic 2--10 keV luminosity.
Table \ref{tab:lum}  lists $L_{Fe}$ (column 4) and 
the resulting values of $L_{AGN,2-10}$ (column 5).
We label this intrinsic 2--10 keV luminosity $L_{AGN,Fe}$ to distinguish
it from the intrinsic luminosity we determine using other techniques.
Typical of Seyfert galaxies, the outcome is 
$L_{AGN,Fe} \sim 10^{42} {\rm \, erg\, s^{-1}}$. 
In all cases, the observed 2--10 keV luminosity 
represents only about
1\% of the intrinsic luminosity 
($L_{obs,2-10}/L_{AGN,Fe}\approx 0.01$.)
Note that the observed luminosities
$L_{obs,0.5-2}$ and $L_{obs,2-10}$ in Table \ref{tab:lum}
are {\em not} corrected for any absorption;
they correspond to the directly-detectable flux of the sources.

\begin{figure}
\includegraphics[width=6in]{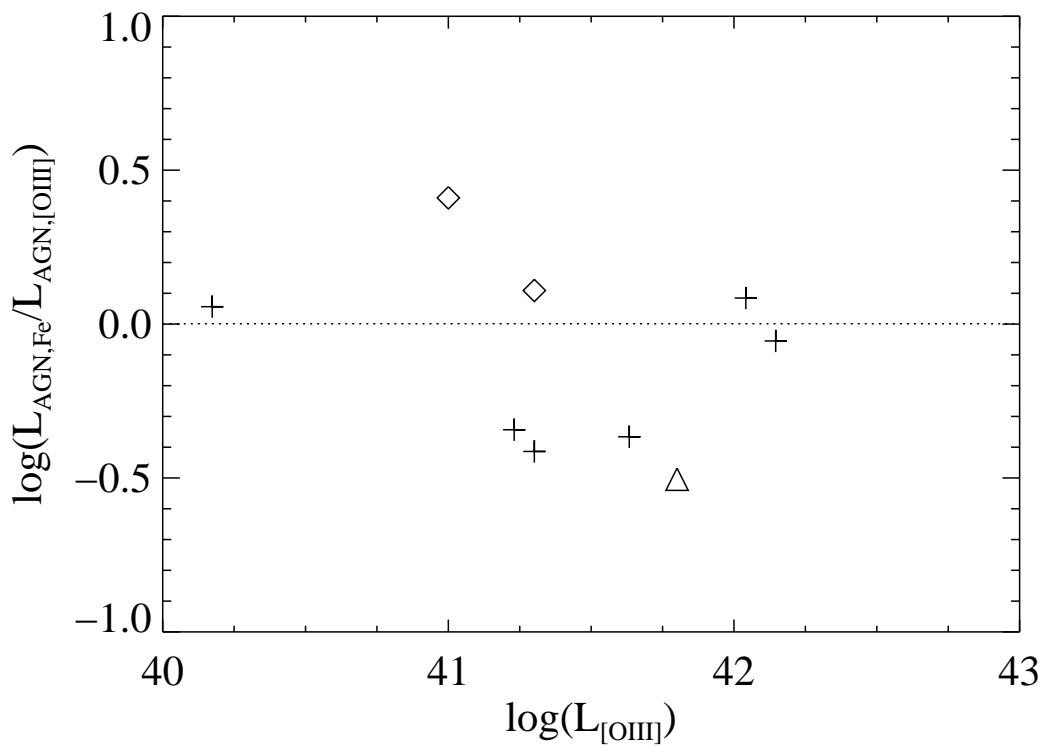} 
\caption{\label{fig:feox} 
Ratio of intrinsic AGN 2--10 keV luminosity determined from the Fe K$\alpha$
line luminosity to intrinsic AGN 2--10 keV luminosity determined from the 
[\ion{O}{3}] luminosity vs. [\ion{O}{3}] luminosity.  
The luminosity is in units of erg s$^{-1}$. 
The sample members
are plotted as crosses, except NGC 1068, which is plotted as a triangle.
Two additional Compton thick galaxies
from the literature that do contain circumnuclear starbursts (NGC 5135 and NGC 7130)
are
plotted as diamonds. 
The line is not a fit to the data and 
shows the good agreement of the two methods for determining intrinsic
luminosity. 
}
\end{figure}

Having the same underlying energy source, $L_{AGN,2-10}$ 
and $L_{[OIII]}$, the luminosity of the [\ion{O}{3}]$\lambda5007$ line, are
correlated.
Our Fe K estimates of the intrinsic luminosity 
are consistent with this  empirical relationship (Table \ref{tab:lum}, column 7).
Specifically, we consider the \citet{Hec05} result
$\log (L_{AGN,2-10}/L_{[OIII]}) = 1.59\pm0.48$
for the optically-selected type 1 AGNs.
(These unobscured nuclei reveal the intrinsic 
relationship, without the significant correction 
for X-ray absorption that even the Compton thin type 2 AGNs
require.)
Figure \ref{fig:feox} illustrates the agreement of the Fe line
and [\ion{O}{3}] estimates of intrinsic AGN luminosity.
In addition to the sample members, which are uncontaminated by
nuclear starbursts, we also plot the published results for two Compton thick
galaxies that do contain starbursts \citep{Lev04,Lev05}.
The dotted line is plotted at a luminosity ratio of one,
not a fit to the data.

We compare the NGC 3393 results with
the {\it BeppoSAX} detection in the 15--220 keV band with the
Phoswich Detector System (PDS) as a direct measurement of the intrinsic AGN.
We use only the high-energy data to measure the intrinsic
AGN continuum level, neglecting the lower-energy measurements,
which include multiple emission sources.
We use the pipeline processed data, fitting the standard grouped spectrum
from 15 to 220 keV.
Both power law and cutoff power law models fit the data, with
$\Gamma \approx 2$.
Extending these models to the \chandra{} bandpass, we
find $F_{AGN,2-10} = 6\pm 2 \times 10^{-11} {\rm \, erg\,cm^{-2} \, s^{-1}},$
or $L_{AGN,2-10} = 2\pm 1 \times 10^{43} {\rm \, erg \, s^{-1}}$,
in reasonable agreement with the Fe line calculation.
\citet{Mai98} and \citet{Gua05} also use the {\it BeppoSAX}
data obtained with multiple instruments over a broader energy range 
and obtain similar results.  Most importantly, their analyses
explicitly show the Compton thick obscuration and characteristic
reflected spectrum that we detect below 8 keV.

\subsection{Origin of the Soft X-rays: Extra-Nuclear Photoionized Gas{\label{subsec:photo}}}

The morphology \citep[\S\ref{sec:obs} and][]{Bia06}
and spectra (\S\ref{sec:spec}) of the soft X-ray emission provide
evidence that we observe
photoionized material. 
The extended features of the X-ray images match those of optical 
images.
In the spectra, we find that 
a limited number of strong lines produce the
observed features, 
in contrast to broad line complexes that would be observed
from thermal plasma at the current spectral resolution,
as we demonstrate quantitatively here. 
Radiation (from the AGN) rather than collisions thus determines
the ionization state of the emitting material.  We therefore describe
this gas as photoionized, although other radiative processes,
including photoexcitation and resonant scattering, likely 
produce some of the observed lines.

\begin{figure}
\includegraphics[width=6in]{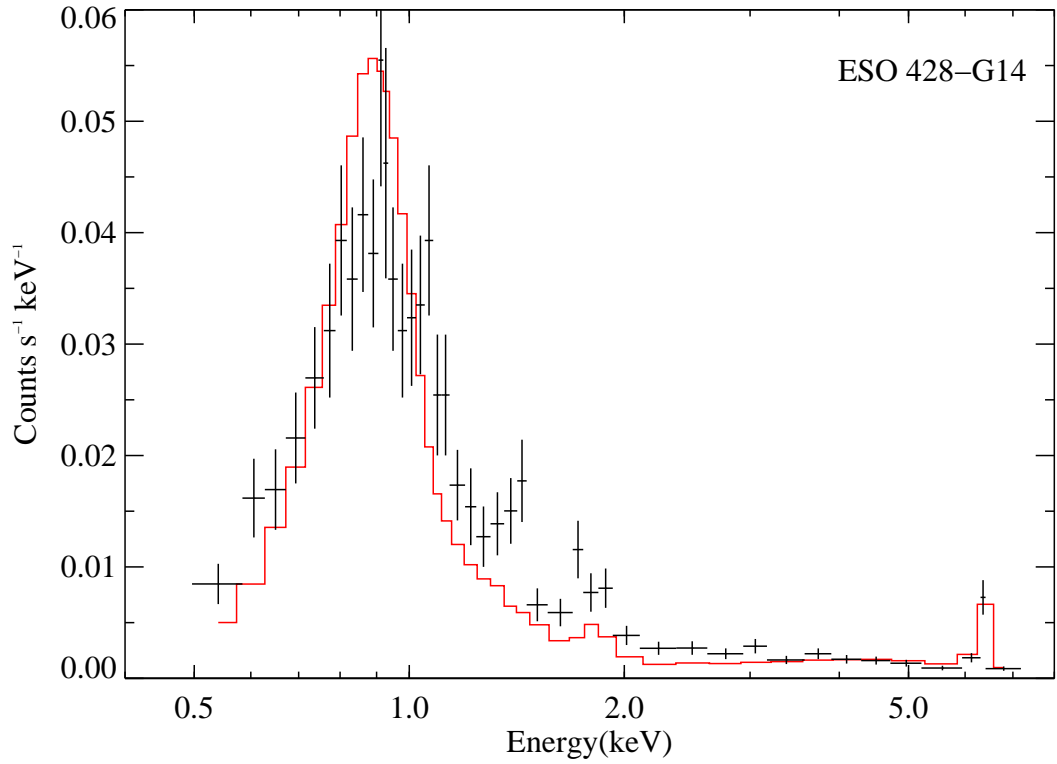} 
\caption{\label{fig:esotherm} 
Nuclear spectrum of \eso{} fit with a thermal model
($kT = 0.7$ keV) in addition to the hard reflection components.
The model fits the soft X-ray emission poorly,
producing a broad emission hump around 0.9 keV,
rather than the observed narrower features.
}
\end{figure}

Figure \ref{fig:esotherm} illustrates the general deficiency
of the thermal models in describing these soft spectra, using \eso{}
as an example.  The best-fitting thermal soft X-ray model 
($kT = 0.70$ keV) fits poorly  ($\chi^2/\nu = 131/42$).  It
results in a broad emission peak around
0.9 keV, and it cannot reproduce the individual narrower peaks
around 0.8, 0.9, and 1.1 keV that are evident at the present resolution.
Two thermal components can fit the spectrum well; with
$kT = 0.63$ and 2.1 keV,  $\chi^2/\nu = 38/40$.
In this case, however, the difficulty would be 
to account for the hot component in a physically reasonable way.
Shock velocities of nearly 1500 km s$^{-1}$ would be required, yet there
is  no evidence for the corresponding broad line widths
in either these spectra or high-resolution spectra of other active
galaxies, where they could certainly be detected. 
Purely thermal models produce similar results in the nuclear spectra of 
the other sample members.  In summary, physically reasonable thermal
models do not quantitatively fit the spectra well.

Thermal emission can be identified and 
measured in CCD X-ray spectra when it is present.
Starburst galaxies, for example, contain large volumes of
hot gas, often in a galactic-scale wind \citep{Hec90}.
Detailed analysis of \chandra{} ACIS observations of the
starburst galaxy NGC 253 clearly shows thermal emission \citep{Str02}.
In this case, a two-temperature model (with $kT = 0.2$ and 0.7 keV)
fits spectra of the diffuse emission well, although it
requires low abundances.

\begin{figure}
\includegraphics[width=6in]{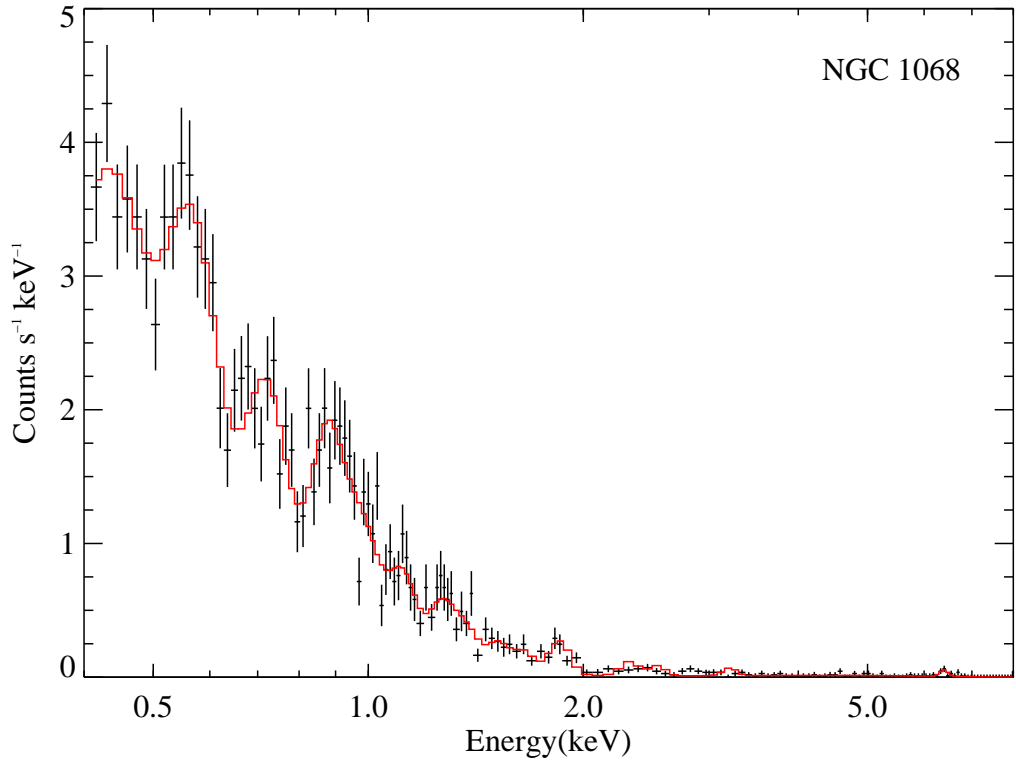} 
\caption{\label{fig:n1068} 
This ACIS spectrum of the nucleus of NGC 1068
shows narrow emission features, 
similar to the the ACIS spectra of the nuclei of 
the other galaxies. 
High spectral resolution observations
of NGC 1068 confirm the dominant role of photoionization in
producing the observed emission.
The observed spectrum ({\it crosses}) 
and total model ({\it red histogram}), uncorrected
for detector sensitivity, are plotted.
}
\end{figure}

NGC 1068 serves as a useful example of photoionized material, 
in which radiative processes produce the soft X-ray line emission.
This brighter galaxy has been successfully observed at high
spectral resolution.   
Measurements of the strength of individual line
transitions reveals photoionization as the
dominant excitation mechanism \citep{Kin02,Ogl03}.
The corresponding spectrum obtained with ACIS 
offers a direct empirical model for the 
other spectra we discuss here.
Figure \ref{fig:n1068} shows the 
spectrum from the
central $1\farcs5$ around the nucleus of NGC 1068.
It is similar to the others,
showing relatively narrow features in the
predominantly soft emission, which we fit well with multiple
Gaussians.

We measure more emission lines significantly in 
this high-quality spectrum compared with the the other sample members,
and we note that much of the flux at energies
above 2 keV is in lines, not continuum (Table \ref{tab:n1068}).
\citet{You01} analyze a spectrum extracted over a  slightly
larger ($1\farcs9$) aperture in these same data.  They show
that the narrow line emission is significant, 
even when strong soft continuum components are included 
in the model.
While interpretation of the physical nature of
X-rays observed at low resolution does depend on the model-fitting
procedure, we emphasize that high quality  CCD spectra are sufficient
to distinguish the thermal or radiative
origin of the emission.

Successful models of X-ray spectra of photoionized line emission
exist,  but 
we do not attempt to apply a more realistic model to these data for two
reasons.
First, although emission line ratios are useful diagnostics of physical conditions,
including identification of photoionized and collisionally-excited plasmas,
we cannot measure any isolated lines in
these data to distinguish among competing model descriptions.
Second,
the models are sensitive primarily to 
the ionization parameter, 
the ratio of radiative flux to gas density, 
given the ionizing spectrum and distribution of illuminated
material.
In the centers of active galaxies, however, no single
ionization parameter is  appropriate.  In such an
inhomogeneous region, individual clumps of gas may have
different densities or be located at different distances
from the ionizing source, resulting in a range of ionization
parameters.  
In fact, high-quality observations 
explicitly show a variety of ionization parameters
within individual  astronomical sources, 
with each line produced predominantly in the region
in which its emissivity is  a maximum \citep[e.g.,][]{Sak99}.

\subsection{Relationship of Soft X-rays and Continuum Emission}

One important result of the spectral fitting is that the soft X-ray
emission of the nuclear and central regions consists almost entirely 
of lines. 
The soft X-ray emission of many Seyfert 2 galaxies has been attributed
to the AGN continuum \citep[e.g.,][]{Rei85,Awa00,Mat04}.  
In this interpretation, although the central engine is generally
obscured, the continuum could be observed through gaps in the 
covering material.
Alternatively,
analogous to the broad emission lines that
are sometimes observed in scattered (polarized) light,
the obscured AGN continuum could be  
scattered into the line of sight. 
These partial covering or scattered
continuum models for the soft X-ray emission were generally plausible
in the past, given the capabilities of existing instruments.
Viewed at low spectral resolution, lines appear blended into
a pseudo-continuum, as others have noted while employing
continuum models as phenomenological descriptions of
their data \citep[e.g.,][]{Tur97}.  
However, the physical interpretation of these
models---in terms of covering fractions and scattering efficiencies, 
for example---will generate misleading conclusions
if fundamentally lines comprise the modeled continuum.

While all  galaxies we analyze here
exhibit strong soft X-ray emission,
it is not the  AGN continuum viewed in reflection or through
patchy obscuration.
Only the central spectrum of \eso{} includes 
such a scattered continuum component in the best-fitting model,
and it accounts for less than one-third
of the 0.5--2 keV flux.  
We consider the possible contribution of a scattered continuum 
component to the soft flux of all the nuclear spectra.  We find strict upper limits 
of 10\% in NGC 1068 and NGC 1386, 20\% in NGC 3393 and NGC 7212,
and 26\% in the nucleus of \eso.
Only the lower-quality spectrum of NGC 5347 does not provide a strong limit,
allowing up to 60\% of the soft flux to be scattered 
continuum if all the emission lines are much weaker than we find in 
best-fitting model.
High-resolution spectroscopy of comparable AGNs 
yields similar results.  
For example, \citet{Kin02} detect no electron-scattered continuum in NGC 1068.
The observed
soft continuum of the Circinus galaxy accounts for less than 30\% 
of the total 0.5--2 keV flux that includes well-measured lines in addition to the
continuum \citep{Sam01}.  In Mrk 3,  
the continuum contributes less than 25\% to 
the observed soft \chandra{} spectrum \citep{Sak00},
although 
\citet{Bia05} indicate that the continuum accounts
for over 50\% of this galaxy's 
flux detected with the Reflection Grating Spectrometer of \xmm{}
up to 1.5 keV, where the line emission is measured well.

In general, the scattered continuum contribution relative to the 
intrinsic emission increases with
decreased clumping factor and increased ionization parameter
\citep{Kro95}.
As these authors demonstrate, 
scattering can also substantially enhance line emission.
This effect is strongest for 
the resonance transitions, 
where the line scattered fraction  can be a few times the
average (continuum) scattered fraction.
Resonance lines are most evident
in the observed spectra, so scattering as well as 
direct recombination likely contributes to these features.

We recognize that partial covering models may provide an accurate physical description
of some other AGNs. 
The spectrum of NGC 4151, for example, has been interpreted in
this context \citep{Wea94,Ogl00}. 
Typically, the ``obscured'' sources in which the 
nucleus appears partially covered at X-ray energies
\citep[e.g.,][]{Wea96,Imm03}
exhibit at least some
broad optical emission lines
(classified as Seyfert types 1.5--1.9).
The broad emission lines, which originate close to the central engine,
are therefore not fully hidden from direct view.
Provided that the covering fraction of the 
the broad emission line region itself is less than 1, 
partial unobscured views of the nuclear continuum are possible.

Obscured active galaxies generally display soft X-ray emission greater than
that of the absorbed  AGN continuum.  
With their continua fully suppressed, these  Compton
thick AGNs show that the luminosity of photoionized line emission 
can be significant, and we suggest that it
may contribute to such ``soft excesses,''
in both obscured and unobscured galaxies.
The 0.5--2 keV luminosity of the concentrated central regions is 
0.3--$11\times 10^{40} {\rm \, erg\, s^{-1}}$, and the
soft luminosity of the entire galaxies ranges from
0.4 to $14\times 10^{40} {\rm \, erg\, s^{-1}}$.
These values are comparable to 
the soft excess luminosity measured in other Seyfert galaxies
\citep[e.g.,][]{Tur97}, or
the total
soft X-ray luminosity of galaxies in which 
all the direct AGN emission is suppressed below 2 keV, i.e., where 
$N_H > 10^{23} \psc$  \citep{Gua05b}.

The soft X-ray emission of the Compton
thick AGNs we investigate here is not related in a simple way to their 
observed or intrinsic hard X-ray emission.
The observed nuclear soft flux ranges from 0.1 to 1 times
the observed hard X-ray flux, and the ratio of soft to hard
flux in the central regions ranges from 0.1 to 0.8.
Assuming the buried AGN generates the soft X-ray lines via photoionization, 
the ratio of observed soft to
intrinsic hard emission indicates a reprocessing efficiency.
This ratio  varies from 
1--$3 \times10^{-3}$, relative to the Fe line estimate
of $L_{AGN,2-10}$.
The production of soft X-rays through photoionization
depends strongly on
the properties  of the surrounding material that may
be illuminated, notably the optical depth and covering fraction. 
In NGC 5347, for example, the lack of dense material along direct lines of sight to
the nucleus is likely responsible for the  low reprocessing efficiency.

\subsection{A Diagnostic Diagram for Obscured AGN}
\begin{figure}
\includegraphics[width=6in]{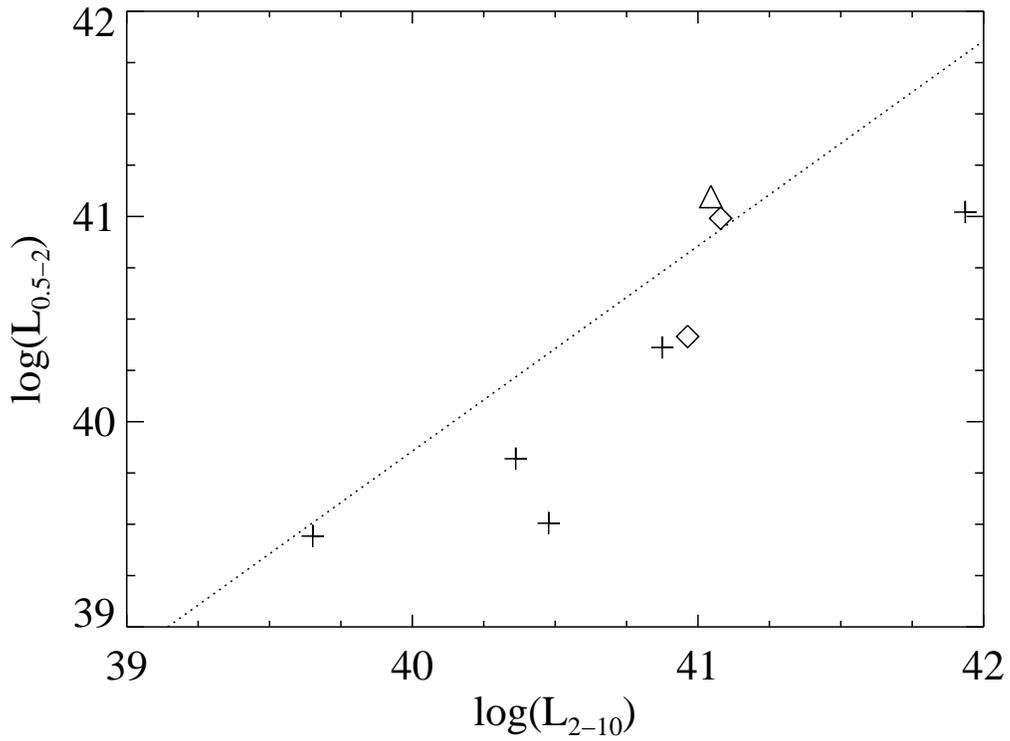} 
\caption{\label{fig:sxhx} 
Observed soft (0.5--2 keV) vs. hard (2--10 keV) luminosity, uncorrected
for absorption, in units of erg s$^{-1}$.  
The line is the relationship for a completely 
{\em unobscured} AGN,
not a fit to the data.  These Compton thick AGNs would appear to be unobscured
and weak based on their hardness ratios alone.
Symbols as in Figure \ref{fig:feox}.
}
\end{figure}

With limited data, such as in surveys, the 
observed ratio of soft and hard fluxes in the form of an
X-ray hardness ratio is commonly employed to
identify an AGN and to determine 
both the degree of obscuration and the intrinsic luminosity.
The hardness ratio 
is an appropriate measurement if the buried AGN dominates the
total X-ray emission and  is detected
directly in the hard X-ray bandpass.  
In such instances, the lack of soft
X-ray emission accurately indicates the absorption while allowing
direct measurement of the true AGN power at higher energies.
In these Compton thick galaxies, however, 
the AGN is not observed directly, and the strong
soft X-ray emission is not the AGN continuum.
For example, defining the  soft band flux ($S$) from 0.5 to 2 keV and the hard band
flux ($H$) from 2 to 8 keV, the hardness ratio is
$(H-S)/(H+S)$.  
{\em Evaluating the hardness ratio in the observed nuclear
or galaxy spectra would incorrectly indicate that these are 
unabsorbed and intrinsically weak AGNs}, having
$N_H< 2\times10^{22} \psc$ and intrinsic
$L_{2-10}<2\times 10^{41} {\rm \, erg \, s^{-1}}$.
Figure \ref{fig:sxhx} illustrates the agreement between
the observed luminosity ratios and that of an {\em unobscured} AGN.

The lack of moderate-energy (2--5 keV) photons
in the presence of strong soft X-rays
distinguishes the heavily obscured 
AGNs from weakly-absorbed yet genuinely faint
sources.
When the data do not allow spectral modeling,
X-ray ``colors''  based on intensity ratios in
at least three bands
would help to identify these sources.
\citet{Gua05b}, for example, consider the effect of redshift and
utilize a very soft (0.2--0.5 keV) band in their analysis.
Alternatively, \citet{Lev05} demonstrate that large column densities can
be identified when gas heated by stellar winds and supernovae
produces the soft X-rays that contaminate
the standard hardness ratio.
We construct a similar grid of AGN models combined
with with varying amounts of photoionization to explore the
X-ray colors in 
soft (0.5--2 keV), medium (2--5 keV), and hard (5--8 keV)
bands.
We model the AGN continuum as either a $\Gamma =1.9$ power law
absorbed by 0--$10^{24} \psc$ or the pure reflection continuum
absorbed by $10^{20}\psc$ (for typical Galactic absorption).
We separately consider including with the reflection model
the flux of an Fe K$\alpha$ line of EW$=1$ keV.
We adopt the empirical soft line spectrum of \eso{} to
represent the photoionized contribution.
We mix the photoionized component with the various AGN
models, scaling it
to account for 0 to 100\% of the total observed counts
with each continuum model.
Finally, for all summed models,
we compute the ratio of counts  in the three bands
\chandra{} would detect.

\begin{figure}
\centerline{\includegraphics[width=4.5in]{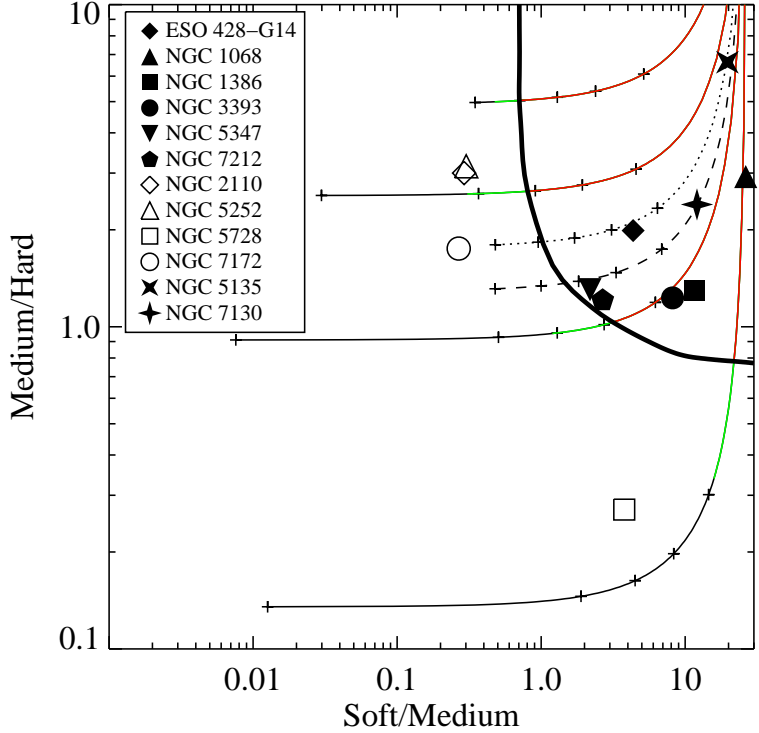}}
\caption{\label{fig:mix} 
X-ray ``color-color'' plot, based on ratios
of counts in the soft (0.3--2 keV), medium (2--5 keV) and hard
(5--8 keV) bands that \chandra{} would observe.
Each curve represents a different AGN continuum model.
Solid curves show fixed AGN obscuration
with $N_H=3\times 10^{22},\ 10^{23},\ 3\times10^{23}$, and $10^{24}\psc$  
(from  top to bottom). 
Alternatively, the AGN 
contribution is calculated from the pure reflection
model,  either excluding ({\it dotted line}) or including ({\it dashed line})
an Fe K$\alpha$ line having  EW $=1$ keV.
An empirical photoionization model  
is mixed with each of these AGN components.
Along each curve, crosses are plotted where the photoionization model
contributes
0, 20, 40, 60, and  80\%
of the total observed counts (from left to right).
Thus, only the leftmost point of each curve
describes a pure AGN model, one without any reflected or scattered soft X-ray 
component.
A constraint on the efficiency with which the
AGN continuum is converted into line emission
excludes higher values of the soft/medium ratio
in the moderate absorption models,
distinguishing them from the reflection models.
{\em Only Compton thick AGNs are found in
the region to the right of the thick solid line.}  Such colors
would 
require excessively high reprocessing efficiency 
when absorption is Compton thin, with 
$L_{photo,0.5-2}/L_{AGN,2-10} \ge 0.03$.
These strongly excluded regions of the
moderately absorbed AGN models are also marked in red.
Green identifies mixes where the reprocessing efficiency is
high, with $0.01 < L_{photo,0.5-2}/L_{AGN,2-10} < 0.03$, although
not strictly requiring Compton thick obscuration. 
The colors of galaxies of this study, the comparison Compton thick
AGNs that also contain starbursts
(both {\it filled symbols}), and several others that
are not Compton thick 
({\it open symbols}) are plotted.
}
\end{figure}

Figure \ref{fig:mix} shows the model intensity 
ratios as a ``color-color'' plot.
The results
 can be ambiguous 
when the photoionized component accounts for 
a significant fraction of the total counts.
In particular, 
three-band ratios alone do not distinguish between the pure reflection
model of the hard X-ray continuum and moderate absorption
($N_H \approx 1$--$3 \times10^{23} \psc$).   
With a significant fraction of  photoionized counts
in these   Compton thin spectra, however, the
photoionization efficiency is unrealistically large.
For $N_H = 10^{23}\psc$, a
 50\% photoionization contribution (by counts)
requires that 6.5\% of the intrinsic AGN's
2--10 keV luminosity emerge 
in lines in  the 0.5--2 keV  band.
For $3\times 10^{23} \psc$, the corresponding luminosity
fraction is 2.1\%.
In contrast, we observe a photoionized
soft X-ray luminosity  $L_{photo,0.5-2}\le 0.005 L_{AGN,2-10}$ 
in the high-quality spectra of the
 Compton thick galaxies of this study.
We therefore suggest imposing an upper limit on the 
allowed photoionization efficiency.
With
the requirement $L_{photo,0.5-2}/L_{AGN,2-10} \le 0.01,$
the photoionized count fraction is restricted to be less
than 14 and 33\%  for  $N_H = 1$ and $3\times 10^{23}\psc$, respectively.  
Equivalently, this constraint rules  out
soft/medium count ratios above 0.2 and 0.9 in
these two Compton thin cases.
We conclusively identify Compton thick galaxies with a strict limit of 
$L_{photo,0.5-2}/L_{AGN,2-10} \ge 0.03$, the area
to the right of the thick solid line in  Figure \ref{fig:mix}.
These excluded Compton thin mixed models are also marked in red.
The model count ratios where
$0.01 < L_{photo,0.5-2}/L_{AGN,2-10} < 0.03$ are plotted
in green.

Figure \ref{fig:mix} also shows the
 observed \chandra{} count ratios of the nuclei
of this study. 
The simplified approach does not reproduce the results of
spectral modeling in detail, but all these AGNs would certainly
be recognized as Compton thick.  For example, NGC 1386 and NGC 3393 are
somewhat weaker in the medium band than the template line emission,
so their colors are offset from those of  the reflection model.
In addition, we plot the observed colors of several 
contrasting Compton thin galaxies,
selected from the available \chandra{} archive and not
classified as the obscured counterparts of narrow-line Seyfert 1
galaxies, which may have unusual soft X-ray properties \citep{Lei99}.
The emission from 
\dataset [ADS/Sa.CXO#obs/04077]{NGC 5728} 
was measured within
a 2\arcsec{} radius aperture.  The bright nuclei of 
\dataset [ADS/Sa.CXO#obs/00883]{NGC 2110}, 
\dataset [ADS/Sa.CXO#obs/04054]{NGC 5252}, 
and 
\dataset [ADS/Sa.CXO#obs/00905]{NGC 7172} 
are strongly piled up, so we calculated the count ratios in the 
resulting readout streak, which remains spectrally accurate.
In the three-color diagram, all these contrasting galaxies would be
classified as
Compton thin, although none of their count ratios
lie close to the values of the single-component
absorbed power law model (at the extreme left of the solid curves).
Finally, we plot the colors of the comparison
Compton thick galaxies that also contain nuclear starbursts.
Because the X-ray emission of the starbursts is almost entirely
soft,  the active nuclei in these composite galaxies remain correctly
classified as Compton thick.
We conclude that  the combination of the three X-ray spectral bands
and a realistic constraint on the efficiency
can identify Compton thick sources without complete spectral modeling.

\section{Continuum Reflection Geometry\label{sec:refl}}

\begin{figure}
\includegraphics[width=6in]{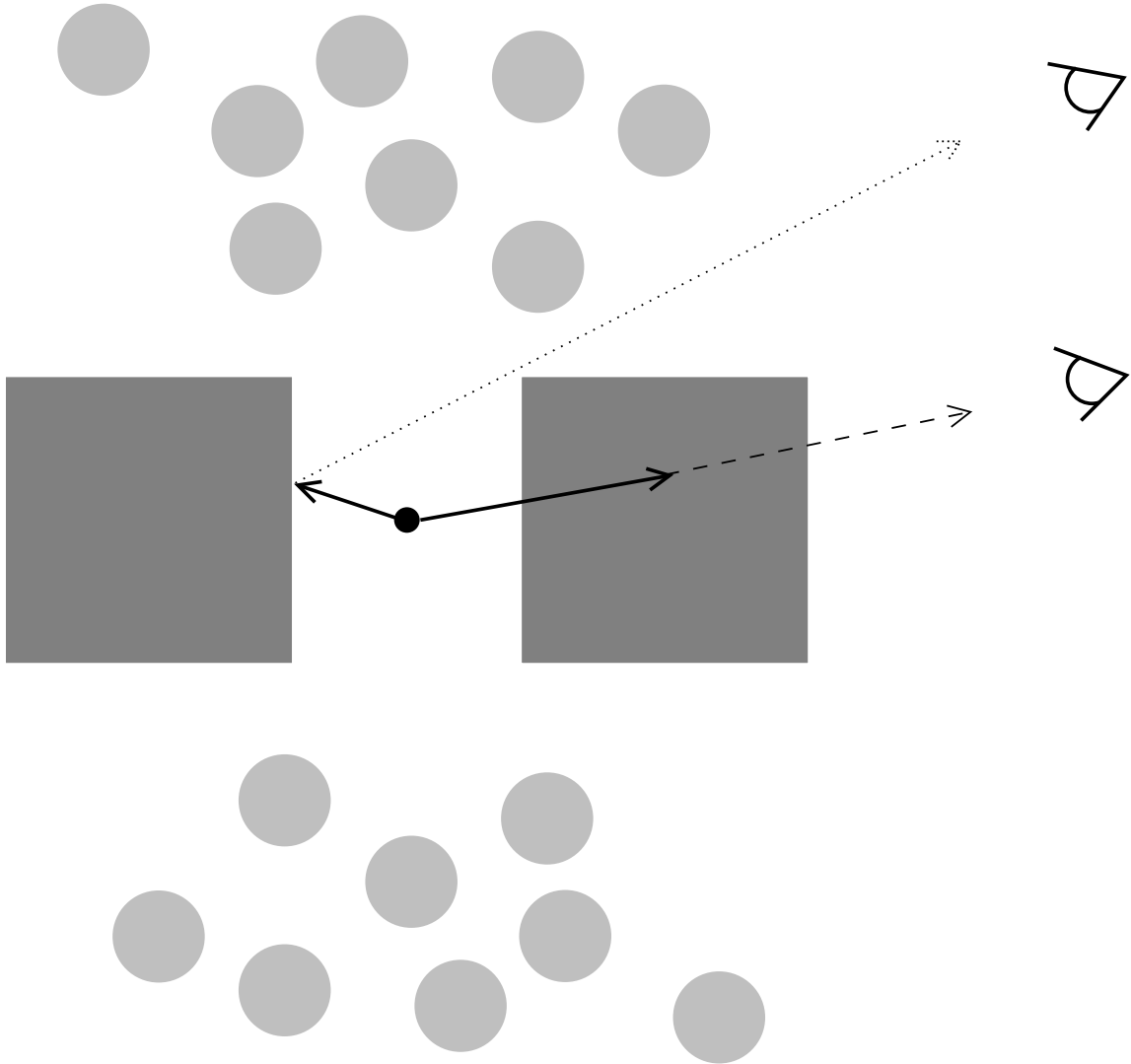} 
\caption{\label{fig:refl} 
Two scenarios to observe the intrinsic AGN
continuum ({\it solid lines}) in reflection,
where an obscuring torus ({\it gray}) blocks
direct views.
For an observer at the right of the cartoon,
the reflecting medium may be the far side of
the torus, and
the reflected light is not further obscured ({\it dotted line}).
Alternatively, the blocking medium may also serve as the
reflecting medium, in which case the reflected spectrum itself
is obscured ({\it dashed line}).
The unobscured continuum reaches
the material of the photoionized emission region ({\it pale gray}).
This figure is {\em not to scale}.  The torus inner radius 
is on the order of 1 pc (the dust sublimation radius), 
while the photoionized emission extends over
hundreds of pc.
}
\end{figure}

The relationship between the observable reflected AGN
continuum and the intrinsic emission is a function of 
geometry.
Figure \ref{fig:refl} schematically illustrates
two scenarios that result in reflection-dominated spectra,
assuming toroidal distribution of obscuring material around
the central engine. 
The intrinsic AGN light (solid lines)  may be
reflected off the far side of the torus 
and then viewed without obscuration (dotted line).
Alternatively, the (near-side) obscuring material
may simultaneously serve as the reflecting medium, in which
case the reflected spectrum (dashed line) is also absorbed, by
an amount that depends on the depth within the material
at which it is produced.
In either case, both the heavily-absorbed intrinsic emission transmitted
through the obscurer and
the reflected component contribute to the net observed spectrum,
but the transmitted component is weak in the Compton thick sources.
For simplicity, the cartoon shows a uniform obscurer, 
although more realistically, the torus may be inhomogeneous.
A clumpy torus results in the same X-ray spectra, 
provided that the total obscuration along the line of sight
is unchanged and---in the first scenario---individual  
clumps remain optically thick to act as reflectors.
(Indeed, rapid X-ray variability observed in other obscured AGNs,
such as NGC 1365 \citep{Ris05}, supports the suggestion
of a non-uniform torus.)
The cartoon also shows the soft X-ray emitting region, 
which has an unimpeded view of the AGN continuum.

Unlike less-obscured examples, in which some portion of the AGN
emission is observed directly, 
recovering the AGN luminosity
from the modeled continuum of these Compton thick galaxies
does not yield certain results.
Interpreting the model fits to the spectra in terms of either of
these scenarios yields a lower limit on 
$L_{AGN,2-10}$.
In the context of  the first scenario,
we assume the
ratio of reflected to intrinsic emission is 1, then
compute the 2--10 keV luminosity of the unseen (buried) continuum source.
The resulting values of $L_{AGN,2-10}$ are listed
in Table \ref{tab:lum} (column 6, labeled $L_{AGN,PEX}$).
Because the relative reflection is in fact less than 1,
with some covering fraction less than 100\% by necessity
allowing direct views of the reflected continuum,
this method gives a lower limit on $L_{AGN,2-10}$.

This estimate of $L_{AGN,2-10}$ is also a lower limit if the second
scenario is applicable.  As in the first situation, the covering
fraction of reflecting material may be less than 1.  The reflected
spectrum may further be obscured by the foreground layers of the
torus.  Viewed through a screen of $N_H= 10^{24}\psc$, for example,
the emerging flux at 8 keV is reduced by a factor of about 5.  Indeed,
the values of $L_{AGN,2-10}$ based on the continuum modeling
are about
an order of magnitude lower than 
the reasonable estimates we derived from the Fe line
luminosities (\S\ref{subsec:feka}), with the exception of NGC 7212. 
Similar reflected and absorbed descriptions of the observed
X-ray emission of obscured AGNs have been proposed previously.
\citet{Awa00}, for example, applied such corrections
to a sample of Seyfert 2s and
increased the estimated intrinsic hard X-ray luminosity by factors
of up to 10 over a single foreground absorber model. 

The reflected soft X-ray emission is detectable in the first case,
while it is absorbed in the second case.
The photoelectric cutoff energy (below which the continuum does not emerge)
is a function of the overlying column density, and a 
cutoff at high energy (indicating large column density) would arise
only in the second scenario.
While the resulting AGN spectra differ,
we cannot discriminate between  these situations in the data.
Line emission obviously dominates the very soft X-ray spectra,
so the presence or absence of the soft reflected continuum is uncertain.
At higher energy, where the spectral cutoff could decidedly identify
the second scenario,
the limited  resolution and sensitivity
prevent distinction of a true continuum from blended lines.

As long as some portion of the far side of the torus is
visible without obscuration, however, this reflected light
dominates the observed hard emission.  This first scenario
is more important even when the
viewing angle is significantly greater than the torus opening
half-angle.  For example, with a viewing angle $i=70\degr$
and $\theta=40\degr$, the 8 keV flux of the
far-side reflection is nearly an order of magnitude greater 
than the emerging near-side reflection.
In the present data, the existence of the photoionized
regions demonstrates that  the covering fractions of the
tori are less than 1, so 
the first scenario best describes the origin of the observed reflection
spectra.

\section{Multiwavelength Comparisons\label{sec:optir}}
While deeply buried, the active nucleus remains the primary energy
source in each of these 
galaxies.
We compare the X-ray results with several other indicators of the luminosity
of these very obscured AGNs at other wavelengths.
Much of the intrinsic luminosity emerges at IR wavelengths,
but stellar emission is a source of confusion.
The extreme ultraviolet (EUV) dominates the intrinsic spectral
energy distribution (SED), and
we probe this regime indirectly through optical recombination lines.
We demonstrate that the intrinsic properties of these AGNs
are similar to their less obscured and unobscured counterparts.
However, their observable characteristics are distinct, 
and therefore offer some identifying signatures.

These Compton thick galaxies generally exhibit 
25- to 60-$\mu$m flux ratios typical of AGNs
\citep[$f_{25}/f_{60} > 0.26$;][]{deG87},
in contrast to cooler, stellar-heated dust.
We calculate the total 8--1000\micron{} infrared luminosity 
from {\it IRAS} fluxes, following the prescription of \citet{San96},
and find $L_{IR}\sim 10^{43} {\rm \,erg\, s^{-1}}$ (Table \ref{tab:lum}). 
The total IR luminosity indicates the bolometric luminosity;
in unobscured quasars, $L_{bol}\approx 5L_{IR}$. 
Ignoring any stellar contributions
and applying this relationship to
these Compton thick sources suggests
$L_{bol} \sim 10^{44}$--$10^{45} {\rm \,erg\, s^{-1}}$ here.

While the ratio of  directly-observed X-ray  to  $L_{IR}$ is very low for AGNs,
with $L_{obs,2-10}/L_{IR} < 10^{-3}$ in all cases,
considering the intrinsic X-ray luminosity, 
$L_{AGN,Fe}/L_{IR} \sim 0.1$.
These latter flux ratios are similar to those measured in 
unobscured sources.  For example, 
in the median radio-quiet quasar spectrum of \citet{Elv94},
$L_{AGN,2-10}/L_{IR} = 0.2$.  \citet{Ris04} derive
the average quasar 
SED in several broad bands,
including corrections for sample selection effects.
This average spectrum has $L_{AGN,2-10}/L_{IR} = 0.05$--0.2, 
depending somewhat on 
the assumed hard X-ray spectral shape and treatment
of the shortest-wavelength IR emission.

\begin{figure}
\includegraphics[width=6in]{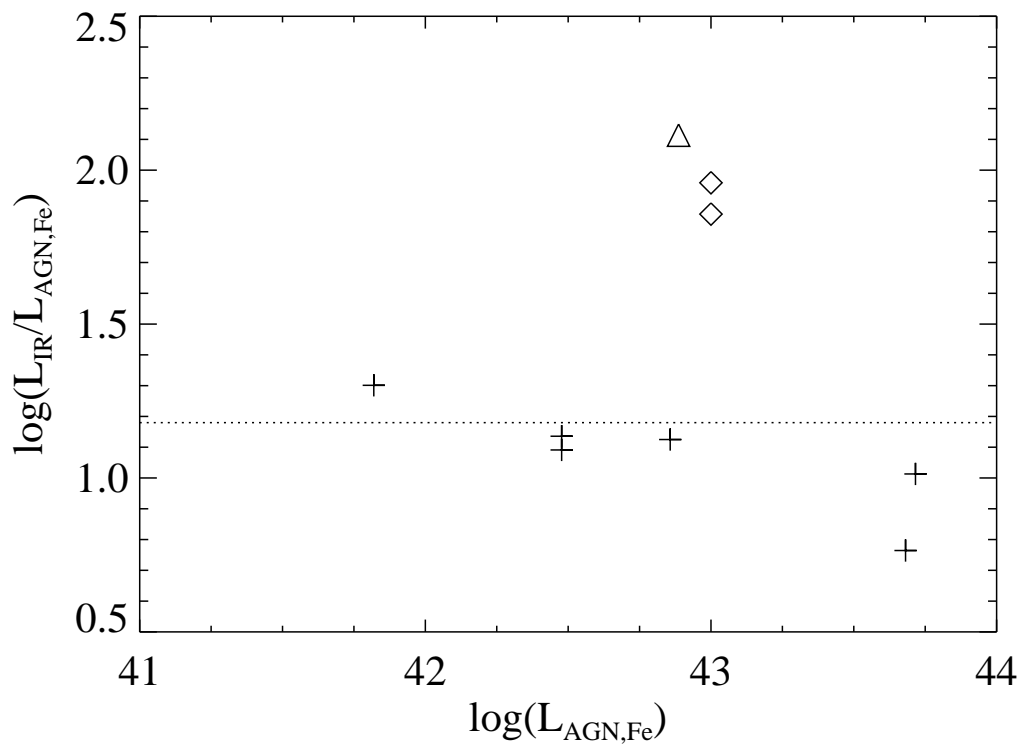} 
\caption{\label{fig:ir} 
Ratio of total IR luminosity to
intrinsic AGN 2--10 keV luminosity determined from the Fe K$\alpha$
line  vs. intrinsic AGN luminosity  determined from the Fe K$\alpha$ line.
The luminosity is in units of erg s$^{-1}$. 
These Compton thick AGNs that do not contain starbursts ({\it crosses}) show
$IR/L_{AGN}$ ratios similar to unobscured AGNs ({\it dotted line}).
The galaxies that do contain strong nuclear starbursts ({\it diamonds})
and NGC 1068 ({\it triangle}), which has a kpc-scale starburst,
exhibit significant excess IR emission.
}
\end{figure}

Neither the composite SEDs nor
our 
AGNs have been corrected for star-heated
dust emission 
at IR wavelengths, which can be important 
within the large {\it IRAS} apertures.
For example, while the kpc-scale star-forming ring of NGC 1068 is easily excluded from
the nuclear X-ray measurements, it does contribute significantly to
the IR measurements.
The comparison of this sample with active galaxies that also 
exhibit significant
star formation demonstrates that 
$L_{IR}$ is not a reliable indicator of intrinsic AGN luminosity
when the stellar populations are unknown.
While the correlations of IR luminosity with
AGN indicators based on [\ion{O}{3}] and
Fe line luminosity are similar for the
non-starburst Compton thick AGNs and unobscured AGNs,
the active galaxies that also contain strong star formation show
IR luminosities that are up to an order of magnitude higher
(Figure \ref{fig:ir}).

\begin{figure}
\includegraphics[width=6in]{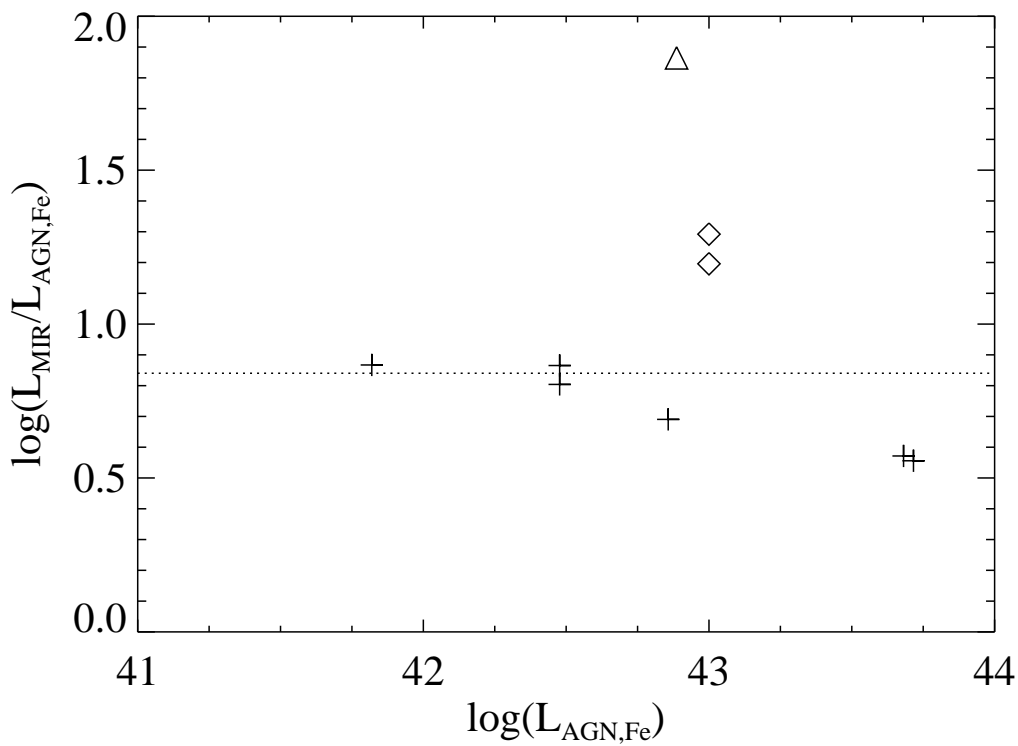} 
\caption{\label{fig:mir} 
Ratio of mid-IR luminosity to
intrinsic AGN 2--10 keV luminosity determined from the Fe K$\alpha$
line  vs. intrinsic AGN luminosity  determined from the Fe K$\alpha$ line.
The luminosity is in units of erg s$^{-1}$. 
Similar to Figure \ref{fig:ir}, the luminosity ratios of
the Compton thick AGNs that do not contain starbursts ({\it crosses}) 
are similar to those of unobscured AGNs ({\it dotted line}).
While the mid-IR excesses of the
starburst galaxies 
are less severe than their total IR excesses, 
mid-IR data alone do not accurately
measure AGN luminosity when stellar contamination is significant.
}
\end{figure}

Mid-infrared emission requires hotter dust, so the stellar contribution
to the integrated luminosity is proportionally smaller in this restricted
wavelength range (8--40\micron).  As Figure \ref{fig:mir} illustrates,
however, even the mid-IR excess of the star-forming galaxies
is significant.  Similar to Figure \ref{fig:ir}, the non-starburst 
Compton thick galaxies exhibit luminosity ratios comparable to
those of unobscured AGNs.  
Although the mid-IR does not always dominate
the total IR luminosity, the two are very well-correlated in non-starburst
galaxies because the underlying energy source---the AGN---is the same.

The large-scale optical emission provides one more method for 
constraining the intrinsic luminosity of these AGNs.
The recombination rate of hydrogen is related to 
the H$\alpha$ line luminosity
$N_{rec} = 7.4\times 10^{11} L_{H\alpha},$
where $L_{H\alpha}$ is measured in erg s$^{-1}$ \citep{Mul94}. 
Assuming every hydrogen-ionizing photon that encounters
circumnuclear material eventually results in recombination,
the ratio of the recombination rate to the
total ionization rate is the covering fraction, $C$, of photoionized material.
We assume a power-law spectrum from the optical to X-ray bandpass,
$L_\nu \propto \nu^{-\alpha}$, with $\alpha$ = 1.5.
Integrating this continuum from the hydrogen ionization edge through
2 keV, the ionizing luminosity 
$L_{ion} \propto L_{H\alpha} C^{-1}$.

We measure $L_{H\alpha}$ and $C$ from the continuum-subtracted
\hst{} image of \eso.
In the resulting  image, we find that line emission
extends over the $3\farcs5 \times 1\farcs1$ region around the
nucleus, for $C = 0.20$.  
We use the WFPC2 exposure time calculator\footnote{http://www.stsci.edu/instruments/wfpc2/Wfpc2\_etc/wfpc2-etc.html}
to convert the observed count rate
to flux for the line spectrum.
We correct for the [\ion{N}{2}] emission which accounts for
about 1/3 of the flux in the
filter bandpass here \citep{Fal96}.  We also
correct for the Galactic extinction 
to yield $F_{H\alpha} = 8.1 \times 10^{-13} {\rm \, erg\, cm^{-2}\, s^{-1}}$.

\citet*{Fer00} report a flux of
$5.4\times 10^{-13} {\rm \, erg\, cm^{-2}\, s^{-1}}$ from H$\alpha$ and
[\ion{N}{2}] in a $6\arcsec \times 8\arcsec$ aperture at the center of NGC 1386.
Because this measurement was made through \hst's linear ramp filter, the
sensitivity to the [\ion{N}{2}] doublet is less than that of H$\alpha$.
We do not correct for the [\ion{N}{2}] contamination here.
We measure the covering fraction of the extended H$\alpha$ emission in the
PC image, finding $C=0.17$.
\citet{Coo00} report a flux of 
$3.67\times 10^{-13} {\rm \, erg\, cm^{-2}\, s^{-1}}$ from H$\alpha$ and
[\ion{N}{2}] in a $7\farcs5 \times 12\farcs4$ area around the 
nucleus of NGC 3393.  Correcting for Galactic extinction and
an estimated 33\% contamination of [\ion{N}{2}],
$F_{H\alpha}=4.4\times10^{-13} {\rm \, erg\,  cm^{-2}\, s^{-1}}$.
We measure the covering fraction of
the {\bf S}-shape evident in the  
\hst{} continuum image, which line emission dominates,
and find $C= 0.26$.
We determine the covering fraction  of NGC 5347 in 
the continuum-subtracted [\ion{O}{3}] image from \hst{} \citep{Schm03}. 
Measuring the opening angle to the northeastern emission,
we find $C = 0.22$.
In narrow-band images, \citet{Gon96} find 
$F_{H\alpha}=7.1\times10^{-13}  {\rm \, erg\,  cm^{-2}\, s^{-1}}$ from
the nucleus and northeastern knot together, after correcting
for [\ion{N}{2}] emission and extinction.

Table \ref{tab:lum} lists $L_{H\alpha}$ and 
the calculated values of $L_{ion}$;
the latter are generally around $10^{43} {\rm \, erg\, s^{-1}}$,
except for the lower-luminosity NGC 1386.
This method of calculating $L_{ion}$ yields a lower limit
on the true value.
The large-scale geometry of the emission-line region 
determines the value of $C$.  
Because the gas is clumpy on small scales, however,
the true covering factor could be substantially smaller, resulting
in larger values of  $L_{ion}$.
Also, while we correct the line fluxes for Galactic extinction,
we do not correct for intrinsic dust absorption.

We associate the X-ray photoionized material with the
optical  narrow line region.
The observed line luminosity 
$L_{H\alpha}$ is typically an order of magnitude
larger than the total observed 0.5--2 keV  luminosity.
The total luminosity of the
narrow line region in optical and UV lines  is then 
a few $\times 10^{42}$ to $10^{43} {\rm \, erg\, s^{-1}}$,
or several orders of magnitude greater, 
even without correcting for intrinsic extinction or covering fraction.
Fundamentally, these luminosity differences demonstrate that the 
emission region is optically thin to X-rays
but optically thick in the 
extreme ultraviolet (EUV), the ionizing continuum of
the optical lines.

Another way to quantify this difference is in terms of the
emission measure, $EM = \int n^2 dl$, where $n$ is density,
integrated along the line of sight.
We treat all of the emission in the X-ray  ``lines'' near 0.56 and 0.91 keV
as \ion{O}{7} and \ion{Ne}{9}, respectively, using the maximum 
line emissivities from the XSTAR photoionization code \citep{Kal01}.
Table \ref{tab:em} lists the emission measures, 
evaluating over the central regions
of \eso, NGC 1386, and NGC 3393, and the nucleus of NGC 5347.
We use
the H$\alpha$ areas and luminosities above to find $EM_{H\alpha}$.
The X-ray emission measures are generally several orders of magnitude
smaller
than the corresponding $EM_{H\alpha}$, even though 
we overestimate all the X-ray line fluxes by 
counting blends of many unresolved lines.
The H$\alpha$ emission measure of NGC 3393
is relatively low,
a consequence of the very large area 
($7\farcs5 \times 12\farcs4$) over which the luminosity
was measured, while the bright $H\alpha$ emission is confined
to a much smaller area ($2\arcsec \times 4\arcsec$).

The viewing geometry of these Seyfert galaxies reveals
the X-ray line emission and optical narrow line regions directly,
and these may be
related to 
the ``warm absorbers''  and UV absorbers observed in unobscured AGNs
\citep{Blu05,Cre03}. 
The present observations and known warm absorbers are
different in several ways, however.
First, 
the X-ray emission measures indicate somewhat lower
column densities (e.g., $N_{OVII} \sim 10^{18}$ rather than $10^{19}\psc$), 
although these are uncertain by
an order of  magnitude, given
the unknown covering fractions.
Second,
the lower ionization
states, such as \ion{O}{7} and \ion{Ne}{9}, 
are the most prominent emission features here.
(High resolution spectra reveal some emission from
fully-ionized species, but it tends to be weaker
and have smaller emission measure; \citealt{Sak00,Kin02}.)
In contrast, fully-ionized species predominate in
the absorption spectra \citep[e.g.,][]{Net03,Yaq03}.
Finally,  the spatial scales of the emission
and absorption appear to be different.
While we measure emission on scales of hundreds of pc,
UV and X-ray absorbers are located within tens of pc of
the AGN \citep[and references therein]{Cre05}.
Although they are not identical,
the absorbing and emitting material need not 
have entirely distinct origins to account for these
observed differences. 
Instead, 
each preferentially samples
a common distribution of properties
differently:
``warm absorption'' spectra  reveal
the smaller scale, rapidly outflowing material, 
while the material 
located farther from the nuclei 
dominates emission.

The multiwavelength data overall answer two important questions:
How is the bolometric luminosity of the AGNs 
distributed across the electromagnetic spectrum, and
at which energies are the active nuclei revealed?
Most of the luminosity clearly emerges in the IR.
The IR spectra and near-constancy
of intrinsic X-ray to IR luminosity ratios demonstrate that 
the buried AGNs produce the observed IR emission in these galaxies
lacking starbursts.
With our calculation of $L_{ion}$, we indirectly recover most but not
all of the 
intrinsic EUV luminosity.
The AGNs themselves
preferentially emit in the EUV, 
and this is the dominant bandpass of quasar
spectral energy distributions \citep{Elv94}, 
although the emergent EUV radiation of all obscured AGNs, 
even the Compton thin variety, is weak. 
Indeed, we  find rough agreement between $L_{ion}$
and $L_{IR}$ as 
estimates of the bolometric luminosity
of each of the buried AGNs.
The sense of the discrepancy between the two wavelength regimes 
($L_{ion} < L_{IR}$)
is expected, for 
the calculated $L_{ion}$ is a lower limit and
the observed $L_{IR}$ may be contaminated by star-heated sources in
the large aperture.
We conclude that
the underlying spectra 
of these Compton thick AGNs are  similar 
to those of unobscured sources, although 
the observable spectra are very different.

The AGN characteristics of Compton thick sources
are not different from other obscured AGNs at optical
wavelengths, so there is no special bias against
finding them in surveys.  The general disadvantages
of optical identification of Seyfert 2s are relevant,
particularly the necessity of obtaining spectra rather than images alone,
and host galaxy dilution of the AGN spectral features, 
which becomes more severe at high redshift \citep{Mor02}.

X-ray surveys  efficiently find active galaxies, despite the
small fraction ($\lesssim 5\%$) of the total luminosity even 
unobscured AGNs emit in this bandpass.
The high sensitivity of the current detectors and the high contrast
with stellar sources (which are proportionately even fainter in X-rays)
particularly contribute to this success.
The Compton thick examples here, however, illustrate the
types of sources that X-ray  surveys can miss.

X-ray observations with even modest spectral resolution
do offer one technique to identify Compton thick
AGNs specifically: the large EW of Fe K$\alpha$, as we and others
have previously suggested \citep{Lev02,Mac04}.  
Measuring the line against the suppressed continuum does not
require complex spectral modeling. 
The line emission is confined to a narrow spectral bandpass,
so the desired signal may easily stand out against
the low background of \chandra{} and \xmm{} within a
limited energy range.  Furthermore, with the EW as a constraint,
the line luminosity is an effective indicator of the intrinsic
AGN luminosity.

Compton thick AGNs may reveal themselves in the IR,
having long-wavelength spectral slopes similar to other types
of AGNs.  An important caveat for IR identification, however,
is that the AGNs do not always dominate the IR emission, even
at mid-IR wavelengths.
Starbursts can be powerful, resulting in IR spectra
that do not indicate the  active nuclei that are present. 
Such stellar contamination is a real concern, because
starbursts are common in active galaxies.

The Compton thick AGNs themselves may be viewed directly 
only  at high energy  ($E> 10$ keV), yet
even then obscuration in excess of $10^{25.5}\psc$ can
reduce the emergent high-energy flux by two orders of magnitude.
Provided the obscuration is not so severe, 
the SWIFT Burst Alert Telescope survey  should  detect
Seyfert galaxies like these and provide direct measurements of
the intrinsic AGN luminosities.  The limiting 15--150 keV flux of the survey will
be 1 mCrab after two years of operation  \citep{Par04}. 
Using the Fe line-determined $L_{AGN,2-10}$ to predict the harder
emission, these sources have flux around 2--3 mCrab.  Other bright
Seyfert galaxies
with intrinsic $L_{AGN,2-10} = 10^{43} {\rm \, erg \, s^{-1}}$ could be
detected to a distance of nearly 80 Mpc.  Truly luminous obscured AGN
having  X-ray luminosities two orders of magnitude greater would be detectable 
to nearly $z = 0.2$.

\section{Conclusions\label{sec:concl}}
Nearby Seyfert galaxies
serve as important examples of
AGNs in general and Compton thick sources specifically.
Because stellar sources do not significantly
contaminate the X-ray emission and the direct AGN
light is blocked, the present observations sensitively reveal the variety
of X-rays that AGNs alone can produce.  
While the underlying continuum may be a simple power law,
the  radiation that emerges is reprocessed, in soft photoionized
lines and a weakened hard continuum.
We find no evidence here that the soft X-ray emission is primarily the
AGN continuum viewed through patchy obscuration or scattered into the
line of sight.
When active nuclei are viewed directly, as in type 1 sources,
the powerful continuum dominates the entire observed spectrum.  
With the direct soft X-rays blocked, however,
even by  smaller column densities 
than in these examples, 
the detectable soft X-rays may be in emission lines.

These spectra clearly show that  line emission, 
not a smooth continuum, composes 
the soft X-ray spectrum.
Although we do not spectrally resolve individual transitions,
the present resolution 
is sufficient to identify photoionization as opposed
to collisional excitation as the
origin of the  line emission.
In all cases, the correlation of the extended soft X-rays with
photoionized line emission at optical wavelengths
supports this interpretation.

Because the observed soft X-ray emission is not the AGN continuum
viewed directly, the commonly-employed hardness ratio does not accurately
measure either
the obscuration or the intrinsic luminosity of these sources.
Instead, we propose a three-``color'' diagnostic
to characterize AGNs---including the Compton thick ones---when 
high-quality data are not available for spectroscopy.

Typical of Compton thick AGNs, even the hard X-rays that Chandra
detects here are reflected, being only about 1\% of
the intrinsic AGN power. 
The clearest X-ray signature of the active nucleus in each of these galaxies
is the large EW Fe K$\alpha$ line emission, with EW $>$ 1 keV in 
all cases.
The EWs are large because the line is measured against a
reflected continuum.  

We determine the intrinsic
luminosity several ways.  The most reliable method in these data
comes from the Fe line luminosity.  The hidden
emission emerges in the IR, and the intrinsic
X-ray/IR flux ratios are typical of unobscured quasars, except where
stellar IR contributions are strong.
Because star-heated dust may be luminous, even at
mid-IR wavelengths, IR data alone do not accurately
measure the AGN luminosity.
The X-ray continuum modeling provides only a lower limit
on the AGN luminosity, independent of the assumed reflection geometry.
The reflection fraction is unknown, and the reflector may itself be
obscured.

\acknowledgements
This research has made use of the NASA/IPAC Extragalactic Database
(NED) which is operated by the Jet Propulsion Laboratory, California
Institute of Technology, under contract with the National Aeronautics
and Space Administration, and it has
made use of data obtained from the High Energy Astrophysics Science Archive Research Center (HEASARC), provided by NASA's Goddard Space Flight Center. 
Some of the data presented in this paper were obtained from the Multimission Archive at the Space Telescope Science Institute (MAST). STScI is operated by the Association of Universities for Research in Astronomy, Inc., under NASA contract NAS5-26555.  
This work was supported by NASA through grant GO4-5117 and 
NSF award AST-0237291.
NAL thanks the GSFC X-ray Astrophysics Laboratory for their hospitality
and the NSFRO for support during part of this work.

{\it Facilities:} \facility{BeppoSAX (PDS)}, \facility{CXO (ACIS)}, \facility{HST (WFPC2)}

\appendix
\section{Additional Sources\label{app:ptsrc}}
In the \chandra{} images, we identify a number of  sources
in addition to the AGNs
within 2\arcmin{} of each galaxy's center.
The sources are detected with a minimum of 3 net counts and non-zero
spatial extent using the CIAO wavdetect algorithm.  
We considered total (0.3--8 keV), low-energy (0.3--2 keV)
and high-energy (4--8 keV) images separately, but the sources
of the latter two images are a subset of sources in the
total-band image.
In all cases, we measure the counts in a 3\arcsec{}
radius aperture after subtracting
a local background.  The source locations and
their net count rates are listed in Table \ref{tab:ptsrc}.
The table also contains 
the projected angular distance of each
source from the nucleus of the Compton thick galaxy in the
field and the corresponding physical scale at the distance
of the central galaxy.

We very crudely estimate the observed fluxes of these sources, assuming
that their spectra are power laws with $\Gamma =1.9$, absorbed
by the Galactic foreground column density.  We find that 
$1\times 10^{-4} {\rm \,count\ s^{-1}}$ corresponds to 
$8\times 10^{-16} {\rm \, erg\, cm^{-2}\, s^{-1}}$.
Individual sources with total X-ray luminosity 
$> 10^{39}{\rm \, erg \, s^{-1}}$ are of interest as ``ultraluminous X-ray''
candidates \citep{Mil04}.  This limit corresponds to
observed count rates of 17, 67, 4, 10, and 0.8 $\times 10^{-4}$ counts s$^{-1}$
in \eso, NGC 1386, NGC 3393, NGC 5347, and NGC 7212, respectively, assuming
the sources are located in the host galaxy.
However, the large apertures in which the source counts
are measured here encompass hundreds of parsecs and
therefore likely include many distinct sources.

These additional sources may be entirely independent of
the galaxies at the centers of these fields.  
The sources around \eso{} and NGC 1386 tend to be soft, however,
and the number of sources in each of these region exceeds 
the source count rates of
blank sky fields.   Applying the results of \citet{Toz01}, for
example, we would expect fewer than about 5 sources
with 0.5--2 keV count rates $> 1\times 10^{-4} {\rm \, count\ s^{-1}}$ 
in each of the $13 \sq\arcmin$ areas we surveyed, while we find 17 and 13 
around \eso{} and NGC 1386, respectively. 
These facts suggest that
many of the sources are indeed associated with (and possibly
located in) these host galaxies. 
Because the total number of sources is small, there is
no significant trend in the 
number of sources as a function of distance from the center of each galaxy.
The number of sources we find near NGC 3393, NGC 5347, and NGC 7212 is
not significantly greater than the detection rates 
expected from blank field observations.
We do measure soft emission from all of the additional sources near NGC 3393
and NGC 7212,  and from half the sources near NGC 5347.

%\documentclass[preprint]{aastex}
%\pagestyle{empty}
%\begin{document}
\begin{deluxetable}{lcclccccc}
\tabletypesize{\scriptsize}
%\rotate
\tablewidth{0pt}
\tablecaption{Observations and Galaxy Data\label{tab:info}}
\tablehead{
\colhead{Galaxy}
&\colhead{Observation Date}
&\colhead{Exposure (ks)}
&\colhead{$N_H$ (cm${\rm^{-2}}$)}
&\colhead{$z$}
&\colhead{Distance (Mpc)}
&\colhead{scale (pc/$\arcsec$)}
&\colhead{RA\tablenotemark{a}}
&\colhead{Decl\tablenotemark{a}}
}
\startdata 
\object[ESO 428-G014]{ESO 428-G14}   &  \dataset[ADS/Sa.CXO#obs/04866]{2003 Dec 26} & 29.6 & $1.2\times 10^{21}$ & 0.0057 & 24.3 & 120 & 07 16 31.18 & -29 19 28.92 \\
\object{NGC 1068}       &  \dataset[ADS/Sa.CXO#obs/00370]{2000 Feb 22} & 1.53 & $2.2\times 10^{20}$ & 0.0038 & 16.2 &  80 & 02 42 40.69 & -00 00 48.11\\
\object{NGC 1386}       &  \dataset[ADS/Sa.CXO#obs/04076]{2003 Nov 19} & 19.6 & $7.7\times 10^{19}$ & 0.0029 & 12.4 &  60 & 03 36 46.18 & -35 59 57.33\\
\object{NGC 3393}       &  \dataset[ADS/Sa.CXO#obs/04868]{2004 Feb 28} & 28.0 & $4.7\times 10^{20}$ & 0.0125 & 53.6 & 260 & 10 48 23.44 & -25 09 43.53 \\ 
\object{NGC 5347}       &  \dataset[ADS/Sa.CXO#obs/04867]{2004 Jun 05} & 33.6 & $1.3\times 10^{20}$ & 0.0078 & 33.4 & 160 & 13 53 17.80 & +33 29 27.46\\
\object{NGC 7212}       &  \dataset[ADS/Sa.CXO#obs/04078]{2003 Jul 22} & 19.9 & $4.5\times 10^{20}$ & 0.0266 & 114 &  550 & 22 07 02.00 & +10 14 01.04 \\
\object{Mrk 3}          &  \nodata & \nodata & $1.2\times 10^{21}$ & 0.0135 & 57.9 & 280  & \nodata & \nodata \\
\enddata
%\tablecomments{}
\tablenotetext{a}{Measured from peak Fe emission.}
\end{deluxetable}
%\end{document}
%%Dist 24.26, 53.57, 33.36 Mpc
%%4piD^2 = 7.07, 34.7, 13.4 x 10^52 cm^2

%\documentclass[preprint]{aastex}
%\pagestyle{empty}
%\begin{document}
\begin{deluxetable}{lccclc}
\tabletypesize{\footnotesize}
%\rotate
\tablewidth{0pt}
\tablecaption{Hard X-ray Spectral Parameters\label{tab:hard}}
\tablehead{
\colhead{Galaxy}
&\colhead{$E$\tablenotemark{a}$_{Fe}$}
&\colhead{$F$\tablenotemark{b}$_{Fe}$}
&\colhead{$EW$\tablenotemark{c}$_{Fe}$}
&\colhead{$A$\tablenotemark{d}$_{AGN}$}
&\colhead{$C/dof$}
}
\startdata 

ESO 428-G14 & $6.35\pm 0.03$ & $7.6\pm 2 $ & $1.6 \pm 0.5$ & $1.7 \pm 0.3$ & $238/270$ \\
NGC 1068\tablenotemark{e} & $6.42\pm 0.05$ & $51^{+30}_{-20}$ & $1.2\pm 0.7$ & $15 \pm 0.5$ & $115/201$ \\
NGC 1386 & $6.39\pm 0.03$ & $5.5^{+2}_{-4}$ & $2.3\pm 1.5$ & $0.8 \pm 0.2$ & $173/270$ \\
NGC 3393 & $6.41\pm 0.07$ & $4.2\pm2$ & $1.4 \pm 0.7$ & $1.0^{+0.3}_{-0.2}$ & $189/270$ \\
NGC 5347 & $6.38\pm 0.03$ & $4.7 \pm 2$ & $1.3\pm 0.5$ & $1.2 \pm 0.2$ & $231/270$ \\
NGC 7212 & $6.39\pm 0.04$ & $8.1\pm 3$ & $1.0\pm 0.4$ & $2.7 \pm 0.5$ & $277/270$ \\
\enddata
\tablecomments{Errors are 90\% confidence limits for one interesting parameter.}
\tablenotetext{a}{Energy of line center in keV.}
\tablenotetext{b}{{L}ine flux in $10^{-6} {\rm \, photons\, cm^{-2}\, s^{-1}}$.}
\tablenotetext{c}{Equivalent width of line in keV.} 
\tablenotetext{d}{Normalization of intrinsic power law in units of
$10^{-3} {\rm\,photons\, keV^{-1}\,cm^{-2}\,s^{-1}}$ at 1 keV.}
\tablenotetext{e}{Fit over 5--8 keV.}
\end{deluxetable}
%\end{document}
%dof = bins - 3

%\documentclass[preprint]{aastex}
%\pagestyle{empty}
%\begin{document}
\begin{deluxetable}{llcll}
\tabletypesize{\footnotesize}
\tablecolumns{5}
%\rotate
\tablewidth{0pt}
\tablecaption{ESO 428-G14 Soft Spectra\label{tab:eso}}
\tablehead{
\multicolumn{2}{c}{Nucleus} &&\multicolumn{2}{c}{Center} \\
\cline{1-2}\cline{4-5}
\colhead{$E$\tablenotemark{a}$_{line}$}
&\colhead{$F$\tablenotemark{b}$_{line}$}
%&\colhead{Identification}
& %blank
&\colhead{$E$\tablenotemark{a}$_{line}$}
&\colhead{$F$\tablenotemark{b}$_{line}$}
%&\colhead{Identification}
}

\startdata 
 $0.61 \pm 0.03$ &   $23\pm 6$    &&  $0.58\pm 0.03$    &  $26 \pm 7$  \\
 $0.77 \pm 0.03$ &   $20\pm 5$    &&  $0.77 \pm 0.02$   &  $23 \pm 6$  \\
 $0.91 \pm 0.02$ &   $18\pm 4$    &&  $0.90 \pm 0.02$  &  $25 \pm 6$   \\
  $1.07\pm 0.02$ &   $11\pm 2$    &&  $1.06 \pm 0.02$   &  $13 \pm 4 $ \\
  $1.23\pm 0.03$ &   $4.3\pm 1$   &&  $ 1.21 \pm 0.04$  &  $4.4 \pm 2$ \\
 $1.42 \pm 0.02$ &   $4.4 \pm 1$  &&  $1.41 \pm 0.03$   &  $4.2 \pm 1$ \\
 $1.73 \pm 0.03$ &   $2.5\pm 0.8$ &&  $1.75 \pm 0.05$   &  $1.8 \pm 1$ \\
  $1.94\pm 0.04$ &   $2.2\pm 0.8$ &&  $1.88 \pm 0.06$   &  $2.0\pm 1$  \\
  $2.40\pm 0.04$ &   $2.6\pm 1 $  &&  $2.36 \pm 0.06$   &  $2.6\pm 1$  \\

%\sidehead{\underline{Additional parameters}}
\cutinhead{Additional Parameters}
 $\Gamma$ &\multicolumn{1}{c}{\nodata} && \multicolumn{2}{c}{1.9f}\\
$A$\tablenotemark{c}$_{\Gamma}$&\multicolumn{1}{c}{\nodata} && \multicolumn{2}{c}{$3.0\pm 0.7$}\\
$\chi^2/dof$ &\multicolumn{1}{c}{$25/26$} &&  \multicolumn{2}{c}{46/49}\\
\enddata
\tablecomments{Errors are 90\% confidence limits for one interesting parameter. Fixed parameters are marked with f.}
\tablenotetext{a}{Energy of line center in keV.}
\tablenotetext{b}{{L}ine flux in $10^{-6} {\rm \, photons\, cm^{-2}\, s^{-1}}$.}
\tablenotetext{c}{Normalization of power law in units of $10^{-5} {\rm\,photons\, keV^{-1}\,cm^{-2}\,s^{-1}}$ at 1 keV.}

\end{deluxetable}
%\end{document}

%\documentclass[preprint]{aastex}
%\pagestyle{empty}
%\begin{document}
\begin{deluxetable}{llcll}
\tabletypesize{\footnotesize}
\tablecolumns{5}
%\rotate
\tablewidth{0pt}
\tablecaption{NGC 3393 Soft Spectra\label{tab:n3}}
\tablehead{
\multicolumn{2}{c}{Nucleus} &&\multicolumn{2}{c}{Center} \\
\cline{1-2}\cline{4-5}
\colhead{$E$\tablenotemark{a}$_{line}$}
&\colhead{$F$\tablenotemark{b}$_{line}$}
%&\colhead{Identification}
& %blank
&\colhead{$E$\tablenotemark{a}$_{line}$}
&\colhead{$F$\tablenotemark{b}$_{line}$}
%&\colhead{Identification}
}

\startdata 
       \nodata    &   \nodata         && $0.40 \pm 0.09$  &  $91\pm 60$    \\
   $0.57\pm 0.03$ &  $26\pm 7$        && $0.56\pm 0.01$    &  $98\pm20$    \\
   $0.70\pm 0.03$ &  $13\pm 5$        && $0.72 \pm 0.02$  &  $33 \pm 6$    \\
   $0.87\pm 0.02$ &  $12 \pm 3$       && $0.88\pm 0.01$   &  $33\pm 5$     \\
%\sidehead{\underline{Additional Parameters}}
\cutinhead{Additional Parameters}
 $kT$ (keV) &  \multicolumn{1}{c}{$1.6\pm 0.3$} && \multicolumn{2}{c}{$1.7 \pm 0.2$}\\
$A$\tablenotemark{c}$_{thermal}$&  \multicolumn{1}{c}{$3.3 \pm 0.7$} && \multicolumn{2}{c}{$8.8\pm 1$}\\
$\chi^2/dof$ &\multicolumn{1}{c}{$22/22$} &&  \multicolumn{2}{c}{$68/53$}\\
\enddata
\tablecomments{Errors are 90\% confidence limits for one interesting parameter.} %Fixed parameters are marked with f.}
\tablenotetext{a}{Energy of line center in keV.}
\tablenotetext{b}{{L}ine flux in $10^{-6} {\rm \, photons\, cm^{-2}\, s^{-1}}$.}
\tablenotetext{c}{Normalization of thermal component in units of $10^{-5} K$, 
where $K=(10^{-14}/(4\pi D^2))\int n_e n_H dV, D$ is the distance to the source (cm), 
$n_e$ is the electron density (${\rm cm^{-3}}$), and $n_H$ is the hydrogen density 
(${\rm cm^{-3}}$).}

\end{deluxetable}
%\end{document}

%\documentclass[preprint]{aastex}
%\pagestyle{empty}
%\begin{document}
\begin{deluxetable}{lll}
\tabletypesize{\footnotesize}
%\rotate
\tablewidth{0pt}
\tablecaption{Observed Flux\label{tab:flux}}
\tablehead{
\colhead{Region}
&\colhead{$F_{0.5-2}$}
&\colhead{$F_{2-10}$}\\
&\colhead{($10^{-13} {\rm \,erg\, cm^{-2}\, s^{-1}}$)}
&\colhead{($10^{-13} {\rm \,erg\, cm^{-2}\, s^{-1}}$)}
}
\startdata 
ESO 428-G14 nucleus &  $0.91\pm 0.2$  &   $3.8\pm 0.5$ \\
ESO 428-G14 center  &  $1.4\pm 0.1$   &   $4.7\pm 0.5$\\
ESO 428-G14 galaxy  &  $1.7\pm 0.3$   &   $6.5\pm 0.7$\\
\\
NGC 1386 nucleus      & $1.1\pm 0.2$  &   $2.1\pm 0.1$ \\
NGC 1386 center       & $1.6\pm 0.2$  &   $2.1\pm 0.4$ \\
NGC 1386 galaxy       &  $2.2\pm 0.2$  &   $3.8\pm 1$ \\
\\
NGC 3393 nucleus      &  $0.73\pm0.07$  &   $2.1 \pm 0.04$\\
NGC 3393 center       &  $2.0\pm 0.2$   &   $2.4\pm 0.07$\\
NGC 3393 galaxy       &  $2.7\pm 0.2$   &   $4.2\pm 0.9$\\
\\
NGC 5347 nucleus      &  $0.24\pm 0.04$ &   $2.2 \pm 0.4$\\
NGC 5347 northeast    &  $0.035\pm 0.01$&     \nodata\\
NGC 5347 galaxy       &  $0.34\pm 0.1$  &   $2.3\pm 0.3$ \\
\\
NGC 7212 nucleus      & $0.43\pm 0.2$  &   $3.9\pm 0.9$ \\
NGC 7212 center       & $0.67\pm 0.3$  &   $5.4\pm 0.6$ \\
NGC 7212 galaxy       & $0.92\pm 0.3$  &    $6.5\pm 0.7$ \\
\enddata
%\tablecomments{}
\end{deluxetable}
%\end{document}

%\documentclass[preprint]{aastex}
%\pagestyle{empty}
%\begin{document}
\begin{deluxetable}{llllllllll}
\tabletypesize{\scriptsize}
\rotate
\tablewidth{0pt}
\tablecaption{Observed and Intrinsic Luminosity\label{tab:lum}}
\tablehead{
\colhead{Galaxy}
&\colhead{$L_{obs,0.5-2}$\tablenotemark{a}}
&\colhead{$L_{obs,2-10}$\tablenotemark{a}}
&\colhead{$L_{Fe}$}
&\colhead{$L_{AGN,Fe}$\tablenotemark{b}}
&\colhead{$L_{AGN,PEX}$\tablenotemark{c}}
&\colhead{$L_{AGN,[OIII]}$\tablenotemark{d}}
&\colhead{$L_{IR}$\tablenotemark{e}}
&\colhead{$L_{H\alpha}$}
&\colhead{$L_{ion}$\tablenotemark{f}}
%\\
%&\colhead{Observed}
%&\colhead{Observed}
%&\colhead{(Fe line)}
%&\colhead{(${\rm erg\, s^{-1}}$)}
%&\colhead{(${\rm erg\, s^{-1}}$)}
}
\startdata 
ESO 428-G14   & $6.6\times10^{39}$ & $2.3\times10^{40}$ & $5.5\times10^{39}$ &
$3.0\times10^{42}$ & $3.6\times10^{41}$ & $7.6\times10^{42}$ &
$4.1\times10^{43}$ & $5.7\times10^{40}$ &
$1.3\times10^{43}$\\
NGC 1068 & $ 1.3\times10^{41}$  & $ 1.1\times10^{41}$  & $ 1.6\times10^{40}$  & $ 7.7\times10^{42}$  & $ 6.3\times10^{41}$  & $ 2.5\times10^{43}$ & $ 1.0\times10^{45}$  & \nodata & \nodata\\
NGC 1386   & $2.0\times10^{39}$  &   $3.9\times10^{39}$  &   $1.1\times10^{39}$  &   $6.6\times10^{41}$  &   $5.2\times10^{40}$  &   $5.8\times10^{41}$ &  $1.3\times10^{43}$ & $1.0\times10^{40}$ & $2.6\times10^{42}$\\
NGC 3393       & $2.3\times10^{40}$ & $7.5\times10^{40}$ & $1.5\times10^{40}$ &
 $7.2\times10^{42}$ & $1.0\times10^{42}$ & $1.7\times10^{43}$ &
$9.6\times10^{43}$ &  $1.5\times10^{41}$ & 
$2.6\times10^{43}$\\
NGC 5347       & $3.2\times10^{39}$ & $3.0\times10^{40}$ & $6.4\times10^{39}$ &
 $3.0\times10^{42}$ & $4.8\times10^{41}$ & $6.5\times10^{42}$ &
$3.7\times10^{43}$ &  $9.5\times10^{40}$ &
$2.0\times10^{43}$\\
NGC 7212   & $ 9.1\times10^{40}$  &   $8.2\times10^{41}$  &   $1.2\times10^{41}$  &   $5.2\times10^{43}$  &   $1.3\times10^{43}$  &   $4.2\times10^{43}$  &   $5.4\times10^{44}$ &    \nodata   &   \nodata \\
Mrk 3    &    \nodata   &   \nodata  &    $2.0\times10^{41}$  &  $ 4.8\times10^{43}$  &   \nodata  &    $5.5\times10^{43}$  &   $2.8\times10^{44}$ &    \nodata   &   \nodata\\
\enddata
\tablecomments{All luminosities are in units of erg s$^{-1}$.}
\tablenotetext{a}{Observed in nuclear spectrum, not corrected for absorption.}
\tablenotetext{b}{Intrinsic 2--10 keV luminosity, determined from Fe line luminosity.}
\tablenotetext{c}{{L}ower limit on intrinsic 2--10 keV luminosity, determined
from PEXRAV spectral model.}
\tablenotetext{d}{Intrinsic 2--10 keV luminosity, determined from [{O} {\sc{iii}}]$\lambda5007$ luminosity.}
\tablenotetext{e}{8--1000$\mu$m bandpass, from {\it IRAS} flux.}
\tablenotetext{f}{{L}ower limit, from optical images.}
\end{deluxetable}
%\end{document}

\begin{deluxetable}{ll}
\tabletypesize{\footnotesize}
\tablecolumns{2}
\tablewidth{0pt}
\tablecaption{NGC 1068 Nucleus\label{tab:n1068}}
\tablehead{
%\multicolumn{2}{c}{Nucleus}\\ 
%\cline{1-2}
\colhead{$E$\tablenotemark{a}$_{line}$}
&\colhead{$F$\tablenotemark{b}$_{line}$}
}
\startdata 

$0.40^{+ 0.01}_{-0.03}$ & $  350^{+180}_{-53}     $ \\
$0.55\pm 0.01         $ & $  180^{+26}_{-14}      $ \\
$0.71^{+ 0.10}_{-0.01}$ & $    71^{+ 7}_{-8}      $ \\
$0.87^{+ 0.14}_{-0.03}$ & $   45^{+ 7}_{-6}       $ \\
$0.99^{+ 0.03}_{-0.04}$ & $    23^{+11}_{-6}      $ \\
$1.12\pm 0.02         $ & $     16\pm 4           $ \\
$1.27\pm 0.30         $ & $    10^{+ 3}_{-4}      $ \\
$1.38^{+ 0.07}_{-0.14}$ & $    6.8^{+ 3}_{-4}     $ \\
$1.53^{+ 0.18}_{-0.10}$ & $    4.7\pm 2           $ \\
$1.66^{+ 0.09}_{-0.04}$ & $    3.6\pm 2           $ \\
$1.88\pm 0.02         $ & $    6.7\pm 2           $ \\
$2.20^{+ 0.06}_{-0.4 }$ & $    3.3\pm 2           $ \\
$2.51^{+ 0.09}_{-0.3} $ & $    6.6^{+5}_{ -2}     $ \\
$3.23^{+ 5.7}_{-0.04} $ & $    2.7^{+3}_{ -4}     $ \\
%$6.42^{+ 0.05}_{-0.06}$ & $    14^{+9}_{ -6}      $ \\
\cutinhead{Additional Parameters}
%$A_{AGN}$\tablenotemark{c} & $4.0\pm 1 $ \\ % pexrav, units 1e-2
$\chi^2/dof$ & \multicolumn{1}{c}{$45/42$}\\
\enddata
\tablecomments{Errors are 90\% confidence limits for one interesting parameter.}
\tablenotetext{a}{Energy of line center in keV.}
\tablenotetext{b}{{L}ine flux in $10^{-5} {\rm \, photons\, cm^{-2}\, s^{-1}}$.}
%\tablenotetext{c}{Normalization of intrinsic power law in units of
%$10^{-2} {\rm\,photons\, keV^{-1}\,cm^{-2}\,s^{-1}}$ at 1 keV.}

\end{deluxetable}

%\documentclass[preprint]{aastex}
%\pagestyle{empty}
%\begin{document}
\begin{deluxetable}{llll}
\tabletypesize{\footnotesize}
%\rotate
\tablewidth{0pt}
\tablecaption{Emission Measures\label{tab:em}}
\tablehead{
\colhead{Galaxy (Region)}
&\colhead{$EM_{OVII}$}
&\colhead{$EM_{NeIX}$}
&\colhead{$EM_{H\alpha}$}\\
&\colhead{(cm$^{-5}$)}
&\colhead{(cm$^{-5}$)}
&\colhead{(cm$^{-5}$)}
}
\startdata 
ESO 428-G14 (center)& $8.0\times10^{20}$ & $2.6\times10^{21}$ & $4.3\times10^{24}$ \\
NGC 1386 (center)   & $1.2\times10^{21}$ & $1.7\times10^{21}$ & $2.1\times10^{23}$ \\
NGC 3393 (center)   & $2.6\times10^{21}$ & $3.0\times10^{21}$ & $9.0\times10^{22}$ \\
NGC 5347 (nucleus)  & $1.8\times10^{21}$ & $1.7\times10^{21}$ & $1.4\times10^{24}$ \\
\enddata
%\tablecomments{}
\end{deluxetable}
%\end{document}

%\documentclass[preprint]{aastex}
%\pagestyle{empty}
%\begin{document}

\begin{deluxetable}{lrrrcrrrrrl}
\tabletypesize{\scriptsize}
\tablecolumns{11}
%\rotate
\tablewidth{0pt}
\tablecaption{Additional Sources\label{tab:ptsrc}}
\tablehead{
&\multicolumn{3}{c}{R.A.}
&&\multicolumn{3}{c}{Decl.}
&\colhead{Count Rate}
&\multicolumn{2}{c}{Distance} \\ 
\cline{10-11} 
&\multicolumn{3}{c}{(J2000)}
&&\multicolumn{3}{c}{(J2000)}
&\colhead{($10^{-4} {\rm \,ct\, s^{-1}}$)}
&\colhead{(\arcsec)}
&\colhead{(kpc)}
}
\startdata
\sidehead{\underline{ESO 428-G14}}
% &  7 &   16 &  23.85  & &     -29 &  21 &   15.38 & $14\pm 3$   & $ 143$ &  16   \\ 
 &  7 &   16 &  25.22  & &     -29 &  20 &   35.19 & $ 2.1\pm 1$ & $ 102$ &  12   \\ 
 &  7 &   16 &  25.45  & &     -29 &  18 &   22.18 & $ 1.8\pm 1$ & $ 100$ &  11   \\ 
% &  7 &   16 &  25.64  & &     -29 &  21 &   25.27 & $ 5.3\pm 2$ & $ 137$ &  16   \\ 
 &  7 &   16 &  26.41  & &     -29 &  21 &    3.43 & $ 2.4\pm 1$ & $ 113$ &  13   \\ 
 &  7 &   16 &  26.75  & &     -29 &  18 &    1.27 & $ 2.1\pm 1$ & $ 105$ &  12   \\ 
 &  7 &   16 &  27.30  & &     -29 &  18 &   26.03 & $ 2.0\pm 1$ & $  81$ &   9.1   \\ 
 &  7 &   16 &  27.66  & &     -29 &  19 &   20.49 & $ 1.1\pm 1$ & $  47$ &   5.3   \\ 
 &  7 &   16 &  29.23  & &     -29 &  21 &   20.45 & $ 2.2\pm 2$ & $ 114$ &  13   \\
 &  7 &   16 &  29.95  & &     -29 &  18 &   20.32 & $26\pm 3$   & $  70$ &   8.0   \\
 &  7 &   16 &  30.07  & &     -29 &  21 &   22.19 & $ 2.2\pm 1$ & $ 114$ &  13   \\
 &  7 &   16 &  30.29  & &     -29 &  21 &   17.81 & $ 3.2\pm 2$ & $ 110$ &  12   \\
 &  7 &   16 &  30.63  & &     -29 &  18 &   35.28 & $ 3.3\pm 2$ & $  54$ &   6.1   \\
 &  7 &   16 &  30.80  & &     -29 &  19 &   35.84 & $ 3.8\pm 2$ & $   9$ &   1.0   \\
 &  7 &   16 &  31.50  & &     -29 &  21 &   12.66 & $ 0.6\pm 1$ & $ 104$ &  12   \\
 &  7 &   16 &  31.85  & &     -29 &  19 &   52.35 & $13\pm 3$   & $  25$ &   2.8   \\
 &  7 &   16 &  31.93  & &     -29 &  19 &   17.77 & $ 4.2\pm 2$ & $  15$ &   1.7   \\
 &  7 &   16 &  32.37  & &     -29 &  17 &   47.56 & $12\pm 2$   & $ 103$ &  12   \\
 &  7 &   16 &  33.01  & &     -29 &  18 &    1.59 & $ 6.2\pm 2$ & $  91$ &  10   \\
 &  7 &   16 &  34.02  & &     -29 &  20 &   22.37 & $ 2.0\pm 2$ & $  65$ &   7.4   \\
 &  7 &   16 &  37.06  & &     -29 &  19 &   28.36 & $ 3.2\pm 2$ & $  77$ &   8.7   \\
 &  7 &   16 &  37.31  & &     -29 &  20 &   48.43 & $ 1.7\pm 1$ & $ 113$ &  13   \\
 &  7 &   16 &  37.33  & &     -29 &  18 &   54.98 & $32\pm 4$   & $  87$ &   9.9   \\
 &  7 &   16 &  37.63  & &     -29 &  20 &   20.76 & $ 2.6\pm 1$ & $  99$ &  11   \\
% &  7 &   16 &  40.97  & &     -29 &  19 &   47.65 & $18\pm 3$   & $ 129$ &  15   \\
\sidehead{\underline{NGC 1386}}
&    3& 36&37.36&& -35&59&10.53&  $  11\pm 2$ & $ 117$ &   7.0\\
&    3& 36&39.17&& -36&00&46.09&  $  11\pm 2$ & $  98$ &   5.9\\
&    3& 36&39.56&& -35&59&30.24&  $   4.9\pm 2$ & $  85$ &   5.1\\
&    3& 36&39.58&& -36&00&56.63&  $   9.7\pm 2$ & $ 100$ &   6.0\\
&    3& 36&43.56&& -36&00&03.90&  $  16\pm 3$ & $  32$ &   1.9\\
&    3& 36&44.18&& -36&00&16.92&  $   3.8\pm 2$ & $  31$ &   1.9\\
&    3& 36&46.31&& -35&59&48.56&  $  21\pm 3$ & $   9$ &   0.5\\
&    3& 36&46.50&& -36&00&05.04&  $  12\pm 3$ & $   9$ &   0.5\\
&    3& 36&46.75&& -35&59&46.56&  $  25\pm 4$ & $  13$ &   0.8\\
&    3& 36&47.32&& -36&00&24.99&  $   5.2\pm 2$ & $  31$ &   1.9\\
&    3& 36&47.61&& -35&59&00.08&  $   3.6\pm 1$ & $  60$ &   3.6\\
&    3& 36&50.22&& -35&59&58.01&  $  19\pm 3$ & $  49$ &   2.9\\
&    3& 36&51.77&& -35&59&08.61&  $   7.6\pm 2$ & $  84$ &   5.0\\

\sidehead{\underline{NGC 3393}}
 &   10 &  48 &  15.67 & &      -25 &     10 &   0.24  & $ 9.2\pm 2$ & $ 107$ &  28  \\
 &   10 &  48 &  17.66 & &      -25 &      9 &   37.24 & $ 0.8\pm 1$ & $  79$ &  20  \\
 &   10 &  48 &  19.13 & &      -25 &     11 &   27.01 & $17\pm 3$   & $ 119$ &  31  \\
 &   10 &  48 &  26.01 & &      -25 &     10 &   23.36 & $13\pm 3$   & $  53$ &  14  \\
 &   10 &  48 &  26.03 & &      -25 &      9 &   15.24 & $ 1.0\pm 1$ & $  45$ &  12  \\
 &   10 &  48 &  27.64 & &      -25 &      8 &   22.87 & $16\pm 3$   & $  99$ &  26  \\
% &   10 &  48 &  29.56 & &      -25 &     11 &   29.46 & $ 9.1\pm 2$ & $ 135$ &  35  \\
 &   10 &  48 &  30.61 & &      -25 &     10 &   6.05  & $54\pm 5$   & $ 100$ &  26  \\
\sidehead{\underline{NGC 5347}}
 &  13 &  53 &   8.33  &&   33 &   29 &  35.45 &   $ 2.4\pm 1$  & $ 119$ &  19  \\  
% &  13 &  53 &   9.34  &&   33 &   31 &  18.43 &   $39\pm 4$    & $ 153$ &  24.8\\    
 &  13 &  53 &   10.69 &&   33 &   29 &  22.41 &   $ 2.1\pm 1$  & $  89$ &  14  \\  
 &  13 &  53 &   17.72 &&   33 &   29 &  8.25  &   $ 2.3\pm 1$  & $  19$ &  3.1 \\  
 &  13 &  53 &   18.33 &&   33 &   27 &  38.84 &   $ 6.0\pm 2$  & $ 109$ &  18  \\  
 &  13 &  53 &   23.85 &&   33 &   30 &  31.41 &   $ 6.2\pm 2$  & $  99$ &  16  \\  
 &  13 &  53 &   24.26 &&   33 &   29 &  36.85 &   $ 1.9\pm 1$  & $  81$ &  13  \\  
 &  13 &  53 &   24.39 &&   33 &   28 &  38.76 &   $ 2.3\pm 1$  & $  96$ &  16  \\  
 &  13 &  53 &   26.38 &&   33 &   30 &  4.86  &   $ 6.7\pm 2$  & $ 114$ &  18  \\  
\sidehead{\underline{NGC 7212}}
&  22&06&54.36&&  10&14&09.02&  $   9.1\pm 2$ & $ 113$ &  63\\
&  22&06&59.05&&  10&15&40.47&  $   7.9\pm 2$ & $ 108$ &  60\\
&  22&06&59.68&&  10&13&13.03&  $   9.3\pm 2$ & $  59$ &  33\\
&  22&07&01.96&&  10&15&16.16&  $   2.3\pm 1$ & $  75$ &  42\\
&  22&07&02.54&&  10&14&16.08&  $   7.9\pm 2$ & $  17$ &   9.5\\
&  22&07&04.23&&  10&13&02.70&  $    11\pm 2$ & $  67$ &  37\\
&  22&07&04.85&&  10&14&22.53&  $   3.8\pm 1$ & $  47$ &  26\\
&  22&07&04.89&&  10&15&08.99&  $   4.0\pm 1$ & $  80$ &  44\\

\enddata
\tablecomments{Units of right ascension are hours, minutes, and seconds, and units of
declination are degrees, arcminutes, and arcseconds.}
\end{deluxetable}
%\end{document}

\clearpage

\end{document}